\documentclass[12pt]{article}
\usepackage[dvips]{graphicx}
\usepackage{graphics}
\usepackage{amssymb}
\usepackage{amsmath,amssymb}
\usepackage{color} 
\usepackage{subfigure}
\usepackage[dvips]{graphicx}
\usepackage{graphics}
\usepackage{dcolumn}
\usepackage{amssymb}
\usepackage{bm}
\usepackage{amsmath}
\usepackage{color}

\def\spose#1{\hbox to 0pt{#1\hss}}

\def\lta{\mathrel{\spose{\lower 3pt\hbox{$\mathchar"218$}}
     \raise 2.0pt\hbox{$\mathchar"13C$}}}
\def\gta{\mathrel{\spose{\lower 3pt\hbox{$\mathchar"218$}}
     \raise 2.0pt\hbox{$\mathchar"13E$}}}
\newcommand{\be}{\begin{equation}}
\newcommand{\en}{\end{equation}}
\newcommand{\bea}{\begin{eqnarray}}
\newcommand{\ena}{\end{eqnarray}}

\begin{document}
\title{General Bianchi IX dynamics in bouncing braneworld cosmology:\\ homoclinic chaos and the BKL conjecture}

\author{
Rodrigo Maier$^{1}$, Ivano Dami\~ao Soares$^{1}$ \\
and Eduardo Valentino Tonini$^{2}$ \\ \\
  $^{1}$ Centro Brasileiro de Pesquisas F\'{\i}sicas -- CBPF, Rio de Janeiro, Brazil\\
  $^{2}$ Instituto de Educa\c{c}\~ao Tecnol\'ogica do Esp\'irito -- IFES, Vit\' oria, Brazil
}


\date{\today}
\maketitle

{\abstract
In the framework of braneworld formalism, we examine the dynamics of a Bianchi IX model with three scale factors
on a 4-dim Lorentzian brane embedded in a 5-dim conformally flat empty bulk with a timelike extra
dimension. The matter content is a pressureless perfect fluid restricted to the brane.
In this scenario Einstein's equations
on the brane reduces to a 6-dim Hamiltonian dynamical system with additional terms --
due to the bulk-brane interaction -- that avoid the singularity and implement nonsingular bounces
in the early phase of the universe. Due to an effective cosmological constant on the brane
the phase space of the model presents two critical points
(a saddle-center-center and a center-center-center) in a finite region of phase space,
and two asymptotic de Sitter critical points at infinity, one acting as an attractor to
late-time acceleration dynamics. The critical points belong to a
2-dim invariant plane; together they organize the dynamics of the phase space.
The center-center-center critical point corresponds to a stable Einstein universe configuration with perpetually oscillatory orbits in its
neighborhood. On the other hand the saddle-center-center engenders in the phase space the topology of stable and unstable
4-dim cylinders $R \times S^3$, where $R$ is a saddle direction and $S^3$ is the center manifold
of unstable periodic orbits, the latter being the nonlinear extension of the center-center sector.
By a proper canonical transformation we are able to separate the three degrees of freedom of the
dynamics into one degree connected with the expansion and/or contraction of the scales of the model,
and two pure rotational degrees of freedom associated with the center manifold $S^3$.
It follows that the typical dynamical flow is an oscillatory
mode about the orbits of the invariant plane. For the stable and unstable cylinders we have
the oscillatory motion about the separatrix towards the bounce, leading to the homoclinic transversal
intersection of the cylinders, as shown numerically in two distinct experiments. We show that
the homoclinic intersection manifold has the topology of $R \times S^2$
consisting of homoclinic orbits biasymptotic to the center manifold $S^3$. This behavior defines a
{\it chaotic saddle} associated with $S^3$, indicating that
the intersection points of the cylinders have the nature of a Cantor set with a compact support $S^2$.
This is an invariant signature of chaos in the model.
We discuss the possible connection between these properties of the dynamics, namely the oscillatory approach to the bounce
together with its chaotic behavior, and analogous features present in the BKL conjecture in general relativity.
}
\\


\section{Introduction}

The general Bianchi IX model has become a paradigm for the behavior of the general relativity dynamics
near the cosmological singularity since the seminal papers of Belinskii, Khalatnikov and Lifshitz (BKL)\cite{bkl1,bkl1o,bkl1oo},
and collaborators\cite{bkl1ooo,bkl2}. They showed that in a Bianchi IX model with three scale factors the approach to the
singularity ($t \rightarrow 0$) is an oscillatory mode, consisting of an infinite sequence of Kasner eras in each of which two
of the scale factors oscillate while the third decreases monotonically. On passing from one era to another
(with decreasing time $t$) the monotonic behavior is transferred to another of the three scale factors.
It was also shown that (i) the length of each era was determined by a sequence of numbers $x_s$, $0 <x_s<1$, $s={\rm integer}$,
each of which arises from the preceding one by the map $x_{s+1}= {\rm the~ fractional~ part~ of~ 1/x_s}$, with
the length of the $s$-th era given by $k_s ={\rm the~ integral~ part~ of~ 1/x_s}$; (ii) this map
leads to spontaneous stochastization in the sequence of eras on approaching the singularity
($t \rightarrow 0$) for arbitrary initial conditions given at $t> 0$.
Due to the involved nonintegrable dynamics the evolution of the model had to be actually
treated in asymptotic regions of arbitrarily small times together with truncations made to guarantee the
validity of the perturbation method, so that ``in the most general case all details of such regime
are not yet fully understood''\cite{bkl1}. In the past four decades the dynamics of
these models has been reexamined in an extensive literature but -- as in the BKL work -- the approach
has been basically twofold: to obtain maps which approximate the dynamics and which exhibit strong stochastic
properties, and the discussion of how well these discrete maps represent the full nonlinear dynamics
of Bianchi IX models in general relativity. From the point of view of the phase space flow the interest
in the chaoticity of Bianchi IX models has been mainly focused on the mixmaster universe (the vacuum Bianchi IX case
with three scale factors), although the question of the behavior (chaotic or not) remained unsettled mainly due
to the absence of an invariant characterization of chaos in the model (standard chaotic indicators as
Liapunov exponents being coordinate dependent and hence questionable). Therefore, along with the Cosmic
Censorship Conjecture, the BKL conjecture is probably one of the major unsolved issues of classical general
relativity connected to the presence of a singularity in the dynamics.
\par Our purpose in the present paper is to examine the dynamics of a 4-dim Bianchi IX model
with three scale factors in the framework of a braneworld formalism (which encompasses general
relativity as a classical low-energy limit). Due to an extra timelike dimension
brane-bulk interaction terms correct general relativity substituting the
singularity by nonsingular bounces in the cosmological dynamics. The dynamics of the approach to
the bounces is extremely complex presenting oscillatory and chaotic features of the BKL-type
but, as we will show, they are amenable to an exact analytical/numerical
treatment so that we may have a more clear picture of what happens in the general relativity limit.
\par
Most of the approaches to the problem of the initial singularity
and to the possible solutions adopted to circumvent
this problem lie in the realm of a quantum theory of
gravitation. In fact we may consider that the initial conditions
of our present expanding Universe were fixed when
the early Universe emerged from a Planckian regime and
started its classical evolution. However, by evolving back
the initial conditions using Einstein's classical equations the
Universe is driven toward a singular point where the classical
regime is no longer valid\cite{wald}. This is an indication that
classical general relativity is not a complete theory and in
this domain quantum processes must be taken into account.
\par
Among several propositions to describe the dynamics in
the semiclassical domain prior to the classical regime are, for instance,
quantum loop cosmology\cite{bojo} and the string based formalism
of D-branes\cite{string}, both of them leading to corrections in
Einstein's equations and encompassing general relativity
as a classical (low energy) limit. In the present paper we adhere to the
so-called braneworld scenario\cite{maeda}-\cite{sahni} based on the string
formalism of D-branes. In this context, extra dimensions
are introduced by a bulk space and all the matter in the
Universe would be trapped on a brane embedded in the bulk with three spatial
dimensions; only gravitons would be allowed to leave the
brane and move into the full bulk\cite{maartens}. At low energies
general relativity is recovered but at high energies significant
changes are introduced in the gravitational dynamics.
Our interest in this framework comes from the fact that it can provide corrections
that are dominant in the neighborhood
of the singularity, resulting in a repulsive force which avoids
it completely and leads the Universe to undergo nonsingular
bounces. Bouncing brane world models were constructed
by Shtanov and Sahni\cite{sahni} based upon a Randall-Sundrum
type action with one extra timelike dimension.
A complete analysis of bouncing brane world dynamics embedded in a five-dimensional
de Sitter spacetime may be found in Refs. \cite{mns,maier}, where both high energy
local corrections as well as nonlocal bulk corrections
are analyzed on a spatially homogeneous brane.
\par
Although spacelike extra dimensions theories have received
more attention in the last decades\cite{maartens}, studies
regarding extra timelike dimensions have been
considered\cite{sak,ya1,chai,ya2}. Albeit presenting some problematic issues\cite{dvali,ynd}
it has been shown\cite{igl} that they might be circumvented by considering a noncompact timelike
extra dimension, which is the case of the model in this paper.
\par In our braneworld scenario we consider a 5-dim de Sitter bulk space with a timelike extra dimension,
and a 4-dim Lorentzian brane with a Bianchi IX geometry with three scale factors.
The matter content of the model is taken as a pressureless perfect fluid (dust)
restricted to the brane and an effective nonvanishing cosmological constant is also considered.
With the above assumptions we show that the Gauss-Codazzi equations, and hence, the modified
field equations on the brane are automatically satisfied. The modified Einstein's equations for
the model have a first integral that can be expressed as a Hamiltonian constraint,
yielding a three degrees of freedom dynamical system in a 6-dim phase space. The additional correction terms due to the
bulk-brane interaction avoid the initial singularity resulting instead to nonsingular bounces
in the model. One of the main features of the phase space is the presence of a saddle-center-center
critical point with an associated center manifold of unstable periodic orbits
having the topology $S^3$. We will show that from the center manifold it emerges stable and unstable
manifolds with the topology of spherical cylinders $R \times S^3$
(constituted actually of bounded oscillatory orbits) which cross each other transversally in
the neighborhood of the bounces. These transversal crossings provide an invariant characterization
of homoclinic chaos in the model.
\par These results are in realm of recent studies in the characterization of homoclinic
chaos for Hamiltonian dynamical systems with $n \geq 2$ degrees of freedom. For $n=2$
the characterization of chaos connected with the presence of homoclinic phenomena in the dynamics
has been the object of an extensive outstanding literature (cf. \cite{berry,guckenheimer,conley,grotta1,grotta2,simo} and references therein).
The dynamics near homoclinic orbits is very complex, with the homoclinic intersection manifold associated
with the presence of the well-known horseshoe structures (cf. \cite{moser,smale,wiggins02} and references therein),
which is an invariant signature of chaos.
Furthermore, invariant Cantor sets associated with a horseshoe construction are connected to chaotic saddles\cite{wiggins04}-\cite{nusse}.
For $n \geq 2$ orbits homoclinic to the center manifold are expected to exist. It has been shown,
for instance, that critical points of the type
saddle-center-...-center induce reaction type dynamics
in the framework of Transition State Theory (see \cite{wiggins1} and references therein).
The existence of such homoclinic orbits has been studied in \cite{wiggins1,wiggins2}.
Although there are no theorems describing the dynamics connected with orbits
homoclinic to $S^3$, it has been shown\cite{cresson} that if there is a
transversal intersection of the stable and unstable manifolds, a
chaotic saddle, and hence an homoclinic trajectory must exist.
An interesting analysis of this feature was given in \cite{wiggins2}, where the authors
provide a computational procedure to detect a chaotic saddle (and thus homoclinic orbits) in the case of Hamiltonian
systems with three degrees of freedom. In the present paper we follow an alternative procedure
to show the presence of homoclinic connections with the center manifold $S^3$.

We organize the paper as follows. In the next section we present a brief introduction
to BraneWorld Theory, deriving the modified field equations on the brane. In Section III
we construct a general Bianchi IX cosmological brane model, with an effective
cosmological constant and the matter content being dust. In Section IV we study the
structure of the phase space, identifying the constants for the linearized motion.
In Section V the dynamics about the saddle-center-center critical point is examined.
Section VI is devoted to a complete analysis of the nonlinear center manifold,
together with the 4-dim stable and unstable cylinders
that emanate from it. Finally in Section VII we study the homoclinic transversal intersections of the cylinders
that gives an invariant characterization of chaos in the dynamics.
Conclusions and future perspectives are presented in the final section.

\section{The field equations}

For sake of completeness we give here a brief introduction to Braneworld Theory, making explicit the specific assumptions
used to obtain the dynamics of the model. We refer to \cite{sahni,maartens} for a more complete and detailed discussion and
our notation closely follows \cite{wald}. We start with a 4-D Lorentzian brane $\Sigma$ with metric
${^{(4)}g}_{ab}$, embedded in a 5-D conformally flat bulk $\cal{M}$ with metric ${^{(5)}g}_{AB}$. Capital Latin indices run from 0
to 4, small Latin indices run from 0 to 3. We regard $\Sigma$ as a common boundary of two pieces
${\cal{M}}_{1}$ and ${\cal{M}}_{2}$ of $\cal{M}$ and ${^{(4)}g}_{ab}$ is the induced geometry on the brane
by the metrics of the two pieces. These metrics should coincide on $\Sigma$ although the extrinsic curvatures
of $\Sigma$ with respect to ${\cal{M}}_{1}$ and ${\cal{M}}_{2}$ can be
different. The action for the theory has the general form

\begin{equation}\label{eqn:eq4r}
\begin{split}
S=\frac{1}{2\kappa^2_{5}}\left\{\int_{M_{1}}\sqrt{-\epsilon~^{(5)}g}\left[^{(5)}R-2\Lambda_{5}
+2\kappa^2_{5}L_{5}\right]d^5x\phantom{^{(5)}g}\right.\\
+\int_{M_{2}}\sqrt{-\epsilon~^{(5)}g}\left[^{(5)}R-2\Lambda_{5}+2\kappa^2_{5}L_{5}\right]d^5x\phantom{^{(5)}g}\\
\left.\phantom{^{(5)}g}+2\epsilon\int_{\Sigma}\sqrt{-^{(4)}g}K_{2}d^4x-2\epsilon\int_{\Sigma}\sqrt{-^{(4)}g}K_{1}
d^4x\right\}\phantom{^{(5)}g}\\
+\frac{1}{2}\int_{\Sigma}\sqrt{-^{(4)}g}\left(\frac{1}{2\kappa^2_{4}}^{(4)}R-2\sigma\right)d^4x\phantom{^{(5)}g}\\
+\int_{\Sigma}\sqrt{-^{(4)}g}L_{4}({^{(4)}g}_{ab},\rho)d^4x\;.\phantom{^{(5)}g}
\end{split}
\end{equation}
In the previous equation, $^{(5)}R$ is the Ricci scalar of the Lorentzian 5-D metric ${^{(5)}g}_{AB}$ on $\cal{M}$, and
$^{(4)}R$ is the scalar curvature of the induced metric ${^{(4)}g}_{ab}$ on $\Sigma$. The parameter $\sigma$ denotes the
brane tension. The unit vector $n^{A}$ is normal to the boundary $\Sigma$ and has norm $\epsilon$. If
$\epsilon=+1$ the signature of the bulk space is $(+,+,-,-,-)$, so that the extra dimension is timelike. The quantity
$K=K_{ab}{^{(4)}g}^{ab}$ is the trace of the symmetric tensor of extrinsic curvature
$K_{ab}=Y_{,a}^{C}~Y_{,b}^{D}~\nabla_{C}n_{D}$, where $Y^{A}(x^a)$ are the embedding functions of $\Sigma$ in
$\cal{M}$\cite{eisenhart}.
While $L_{4}({^{(4)}g}_{ab},\rho)$ represents the Lagrangian density of the perfect fluid\cite{taub} (with equation of state
$p=\alpha\rho$), whose dynamics is restricted to the brane $\Sigma$, $L_{5}$ denotes the Lagrangian of matter in the
bulk. All integrations over the bulk and the brane are taken with the natural volume elements
$\sqrt{-\epsilon~{^{(5)}}g}~d^{5}x$ and $\sqrt{-{^{(4)}}g}~d^{4}x$ respectively. Einstein constants in five- and
four-dimensions are indicated with $\kappa^2_{5}$ and $\kappa^2_{4}\equiv8\pi G_N$, respectively ($G_N$ being the Newton's constant on the brane). Throughout this section we use
natural units with $\hbar=c=1$.

\par Variations that leave the induced metric on $\Sigma$ intact, furnish the equations

\begin{equation}\label{eqn:eq6r}
^{(5)}G_{AB}+\Lambda_5~{^{(5)}}g_{AB}=\kappa^2_5~{^{(5)}}T_{AB}\;.
\end{equation}
Considering arbitrary variations of ${^{(5)}g}_{AB}$ and taking into account Eq.~(\ref{eqn:eq6r}), we obtain

\begin{equation}\label{eqn:eq7r}
^{(4)}G_{ab}+\epsilon\frac{\kappa^2_4}{\kappa^2_5}\left(S^{(1)}_{ab}+S^{(2)}_{ab}\right)
=\kappa^2_4\left(\tau_{ab}-\sigma g_{ab}\right)\;,
\end{equation}
where $S_{ab}\equiv K_{ab}-K{^{(4)}g}_{ab}$, and $\tau_{ab}$ is the energy momentum tensor on the brane. In the limit
$\kappa^2_4\gg\kappa^2_5$,
Eq.~(\ref{eqn:eq7r}) reduces to
the Israel-Darmois junction condition{\bf s}\cite{israel}

\begin{equation}\label{eqn:eq9r}
\left(S^{(1)}_{ab}+S^{(2)}_{ab}\right)=\epsilon\kappa^2_5\left(\tau_{ab}-\sigma {^{(4)}g}_{ab}\right)\;.
\end{equation}

Imposing the $Z_2$-symmetry\cite{maartens} and using the junction conditions (Eq.~\ref{eqn:eq9r}), we determine the
extrinsic curvature on the brane,
\begin{equation}\label{eqn:eq10r}
K_{ab}=\frac{\epsilon}{2} \kappa^2_5 \left[\left(\tau_{ab}-\frac{1}{3}\tau {^{(4)}g}_{ab}\right)+\frac{\sigma}{3}
{^{(4)}g}_{ab}\right]\;.
\end{equation}

\par Now using Gauss equation

\begin{equation}\label{eqn:gauss}
\begin{split}
^{(4)}R_{abcd}=^{(5)}R_{MNRS} Y^{M}_{,a} Y^{N}_{,b} Y^{R}_{,c} Y^{S}_{,d}-\\
\epsilon \left(K_{ac}K_{bd}-K_{ad}K_{bc} \right)\;,
\end{split}
\end{equation}
together with Eqs.~(\ref{eqn:eq6r}) and (\ref{eqn:eq10r}) we obtain the induced field equations on the brane
\begin{equation}\label{eqn:eq1.2.13}
\begin{split}
^{(4)}G_{ab}+\Lambda_{4}{^{(4)}g}_{ab}=8\pi G_{N}\tau_{ab}-\epsilon\kappa^4_{5}\Pi_{ab}+\\
\epsilon{E}_{ab}+\epsilon F_{ab}\;.
\end{split}
\end{equation}
In the above $E_{ab}={^{(5)}}C_{ABCD}n^A Y^{B}_{,a}n^{C} Y^{D}_{,b}$ is the projection of the 5-D Weyl tensor, and we have defined

\begin{eqnarray}\label{eqn:eq1.2.14}
\Lambda_{4}
& = & \frac{1}{2}\kappa^2_{5}\left(\frac{\Lambda_{5}}{\kappa^2_5}-\frac{1}{6}\epsilon\kappa^2_{5}\sigma^2\right)\;,\\
G_{N} & = & \epsilon\frac{\kappa^4_{5}\sigma}{48\pi}\;,\\
\nonumber
\Pi_{ab} & = & -\frac{1}{4}\tau_{a}^{c}\tau_{bc} +\frac{1}{12}\tau\tau_{ab}
+\frac{1}{8}{^{(4)}g}_{ab}\tau^{cd}\tau_{cd}-\\
& & \frac{1}{24}\tau^2{^{(4)}g}_{ab}\;,\\
\nonumber
F_{ab} & = & \frac{2}{3}\kappa^2_{5}\Big\{\epsilon ~{^{(5)}T}_{BD} Y^B_{,a} Y^D_{,b}-\\
& & \Big[{^{(5)}T}_{BD} n^{B} n^{D}+\frac{1}{4}\epsilon~{^{(5)}T}\Big]{^{(4)}g}_{ab}\Big\}\;,
\end{eqnarray}
Here we stress that the effective 4-dim cosmological constant can be set to zero
in the present case of an extra timelike dimension by properly fixing the bulk cosmological constant as
$\Lambda_{5}=\frac{1}{6}\kappa_{5}^{4} ~\sigma^2$.
It is important to notice that for a 4-dim brane embedded in a conformally flat empty bulk we have the absence of the Weyl
conformal tensor projection $E_{ab}$, and of $F_{ab}$ in Eq.~(\ref{eqn:eq1.2.13}).

On the other hand, Codazzi's equations imply that
\begin{equation}\label{eqn:eq1.2.17}
\nabla_{a} K-\nabla_{b}K^{b}_{a}=\frac{1}{2}\epsilon \kappa^2_{5}\nabla_{b}\tau^{b}_{a}\;.
\end{equation}
By imposing that $\nabla_{b}\tau^{b}_{a}=0$, the Codazzi conditions read
\begin{equation}\label{eqn:eq1.2.18}
\nabla^{a}{E}_{ab}=\kappa^4_{5}\nabla^{a}\Pi_{ab}+\nabla^{a}F_{ab}.
\end{equation}
where $\nabla_a$ is the covariant derivative with respect to the induced metric ${^{(4)}g}_{ab}$. Eqs.~(\ref{eqn:eq1.2.13}) and
(\ref{eqn:eq1.2.18}) are the dynamical equations of the gravitational field on the brane. In the following section
we drop the index {\small $(4)$} in the geometrical quantities on the brane.

\section{The model}

Let us consider a Bianchi IX spatially homogeneous geometry on the four-dimensional brane embedded
in a five-dimensional, conformally flat and empty bulk ($E_{ab}=0=F_{ab}$) with a timelike extra dimension ($\epsilon=1$).
In comoving coordinates on the brane, the line element can be expressed as
\begin{eqnarray}
\label{eq2}
ds^2=dt^2 - (\theta^{1})^2 - (\theta^{2})^2- (\theta^{3})^2,
\end{eqnarray}
where $t$ is the cosmological time and
\begin{eqnarray}
\label{eq3}
\theta^{1}=M(t)~ \omega^1,~~ \theta^{2}=N(t)~ \omega^2,~~ \theta^{3}=R(t)~ \omega^3.
\end{eqnarray}
Here $M(t)$, $N(t)$ and $R(t)$ are the scale factors of the model and the $\omega^{i}$ ($i=1,2,3$) are
Bianchi-type IX 1-forms satisfying
\begin{eqnarray}
\label{eq4}
d\omega^{i}= \frac{1}{2}\epsilon^{ijk} \omega^{j} \wedge\omega^{k},
\end{eqnarray}
where $d$ denotes the exterior derivative.
\par The matter content of the model is assumed to be dust,
whose energy density $\rho$ is measured by the comoving observers
with $4$-velocity $u^{a}=\delta_{0}^{a}$.
By imposing that the energy-momentum tensor of dust,
\begin{eqnarray}
\label{eq5}
\tau^{ab}=\rho u^{a} u^{b},
\end{eqnarray}
is conserved separately, namely $\nabla_{a} \tau^{ab}=0$, we obtain
\begin{eqnarray}
\label{eq6}
\rho= \frac{C_0}{M N R},
\end{eqnarray}
where $C_0$ is a constant of motion connected to the dust energy.
The components of tensor $\Pi_{ab}$ are given by
\begin{eqnarray}
\label{eq7}
\Pi_{ab}=\frac{1}{12}\rho^2 ~ g_{ab},
\end{eqnarray}
so that Codazzi's equations,
\begin{eqnarray}
\label{eq6}
\nabla_{a}\Pi^{ab}=0,
\end{eqnarray}
are identically satisfied. Therefore, Eq. (\ref{eqn:eq1.2.13}) reduces to
\begin{equation}\label{eqnG}
G_{ab}+\Lambda~{g}_{ab}=8\pi G_{N}\tau_{ab}-\kappa^4_{5}\Pi_{ab},
\end{equation}
which are the modified Einstein's field equations for the model.
As the Gauss-Codazzi equations are automatically satisfied via (\ref{eq6}) and (\ref{eqnG}),
the assumption of a conformally flat empty bulk is consistent. We also see that as $\epsilon=1$
(a timelike extra dimension) the term $\Pi_{ab}$ in (\ref{eqnG})
acts as a potential barrier to the dynamics avoiding the singularity.
\par
In terms of the metric functions (\ref{eq2}) equations (\ref{eqnG}) correspond to the modified Friedmann's equations
of the model, having a first integral that can be expressed as the Hamiltonian constraint
%
\begin{eqnarray}
\label{eq4}
\nonumber
H&=&\frac{1}{8}\Big(-\frac{M}{NR}p^2_M-\frac{N}{MR}p^2_N-\frac{R}{MN}p^2_R+\frac{2}{M}p_Np_R
+\frac{2}{N}p_M p_R+\frac{2}{R}p_M p_N\Big)\\
\nonumber
&+&\frac{1}{2MNR}[M^4+N^4+R^4
-(M^2-N^2)^2-(R^2-M^2)^2-(R^2-N^2)^2]\\
&-& 2\Lambda MNR-2E_0+\kappa^2\frac{E_0^2}{MNR}=0~~
\end{eqnarray}
%
where $p_M$, $p_N$ and $p_R$ are the momenta canonically conjugate to
$M$, $N$ and $R$, respectively. $E_0\equiv 8\pi G_N C_0$ and
$\kappa^2\equiv (8\pi G_N)^{-1}|\sigma|^{-1}$.
From Hamilton's equations we obtain the following dynamical system
%
\begin{eqnarray*}
\label{eqH}
\nonumber
\dot{M}&=&\frac{\partial H}{\partial p_M}=\frac{1}{4}\Big(\frac{p_N}{R}+\frac{p_R}{N}-\frac{M}{NR}p_M\Big),~~~~~~~~~~~~~~~~~~\\
\nonumber
\dot{N}&=&\frac{\partial H}{\partial p_N}=\frac{1}{4}\Big(\frac{p_M}{R}+\frac{p_R}{M}-\frac{N}{MR}p_N\Big),~~~~~~~~~~~~~~~~~~\\
\dot{R}&=&\frac{\partial H}{\partial p_R}=\frac{1}{4}\Big(\frac{p_M}{N}+\frac{p_N}{M}-\frac{C}{MN}p_R\Big),~~~~~~~~~~~~~~~~~~\\
\nonumber
\dot{p}_M&=&-\frac{\partial H}{\partial M}=\frac{1}{8}\Big(\frac{p^2_M}{NR}-\frac{N}{M^2R}p^2_N-\frac{R}{M^2N}p^2_R
+\frac{2}{M^2}p_N p_R\Big)\\
\nonumber
&&+\frac{1}{2M^2NR}[M^4+N^4+R^4-(R^2-N^2)^2
-(R^2-M^2)^2-(M^2-N^2)^2]\\
\nonumber
&&+2\Lambda NR -\frac{2}{MNR}[M^3+M(R^2-M^2)-M(M^2-N^2)]+\frac{\kappa^2E^2_0}{M^2NR},\\
\dot{p}_N&=&-\frac{\partial H}{\partial N}=\frac{1}{8}\Big(\frac{p^2_N}{MR}-\frac{M}{N^2R}p^2_M-\frac{R}{MN^2}p^2_R
+\frac{2}{N^2}p_M p_R\Big)\\
\nonumber
&&+\frac{1}{2MN^2R}[M^4+N^4+R^4-(R^2-N^2)^2
-(R^2-M^2)^2-(M^2-N^2)^2]\\
&&+2\Lambda MR
-\frac{2}{MNR}[N^3+N(R^2-N^2)-N(N^2-M^2)]+\frac{\kappa^2E^2_0}{MN^2R},\\
\end{eqnarray*}
\begin{eqnarray}
\label{eqH-vi}
\nonumber
\dot{p}_R&=&-\frac{\partial H}{\partial R}=\frac{1}{8}\Big(\frac{p^2_R}{MN}-\frac{M}{NR^2}p^2_M-\frac{N}{MR^2}p^2_N
+\frac{2}{R^2}p_M p_N\Big)\\
&&+\frac{1}{2MNR^2}[M^4+N^4+R^4-(R^2-N^2)^2
\nonumber
-(R^2-M^2)^2-(M^2-N^2)^2]\\
&&+2\Lambda MN
-\frac{2}{MNR}[R^3+R(M^2-R^2)-R(R^2-N^2)]+\frac{\kappa^2E^2_0}{MNR^2}~.
\end{eqnarray}
%
Equations (\ref{eq4}) and (\ref{eqH}) are equivalent to the modified field equations (\ref{eqnG}).
\section{The structure of the phase space\label{sectionIV}}

In this section we will examine the basic structures that organize the
dynamics of the system in the phase space. The first of these are the
set of critical points of the system given, from Eqs. (\ref{eqH})),
by $M=N=R=M_0$ and $p_M=p_N=p_R=0$, where $M_0$ satisfies the equation
\begin{eqnarray}
\label{eqCr}
M_0^6 - \frac{M_0^4}{4 \Lambda}+\kappa^2~\frac{E_{\rm cr}^2}{2 \Lambda}=0.
\end{eqnarray}
We can observe that the critical points, determined by the positive real roots of (\ref{eqCr}),
depend on their respective critical energy appearing in the third term of the left-hand-side of the equation,
as a consequence of the bulk-brane interaction.
\par We must also consider the further relation
\begin{eqnarray}
\label{eqHC}
\frac{3}{2}M_0 +\kappa^2~\frac{E_{\rm cr}^2}{M_0^3}- 2 \Lambda M_0^3 - 2 E_{\rm cr}=0,
\end{eqnarray}
obtained by evaluating the Hamiltonian constraint (\ref{eq4}) at the critical points.
Solving (\ref{eqHC}) for $E_{\rm cr}$ we will restrict ourselves to the root
\begin{eqnarray}
\label{eqHCroot}
E_{\rm cr}= \frac{M_0^3}{\kappa^2}\Big( 1- \sqrt{1-\frac{3 \kappa^2}{2 M_0^2}+2 \kappa^2 \Lambda}~~\Big)
\end{eqnarray}
which yields the correct result in the general relativity limit\cite{ozorio}
($\kappa^2 \rightarrow 0$ or equivalently $|\sigma| \rightarrow \infty$).
Combining Eqs. (\ref{eqCr}) and (\ref{eqHCroot}) we obtain for the critical points
the two real positive solutions
%
\begin{eqnarray}
\label{eqM012}
M_{0_{1,2}}= \frac{\kappa~ (3 \pm \sqrt{1-16 \kappa^2 \Lambda}~)}{2(1+2 \kappa^2 \Lambda) \sqrt{(1-4 \kappa^2 \Lambda \pm \sqrt{1-16 \kappa^2 \Lambda}~)/(1+2 \kappa^2 \Lambda)}}
\end{eqnarray}
%
with $M_{0_{1}} \leq M_{0_{2}}$. The equality occurs for $\Lambda=1/16 \kappa^2$, the case of just one critical point; for $\Lambda > 1/16 \kappa^2$
no critical points exist. In the following we are going to restrict ourselves to the case $\Lambda<1/16 \kappa^2$. As we will see, this condition is
necessary for the presence of homoclinic orbits that establish the chaotic behavior of the dynamics.
The respective energies associated with the critical points are obtained by substituting $M_{0_{1,2}}$ in (\ref{eqHCroot})
yielding
\begin{eqnarray}
\label{eqEcr1}
E_{{\rm cr}_{(1,2)}}= \frac{\kappa~ (3 \pm \sqrt{1-16 \kappa^2 \Lambda}~)^2}{8~ (1+2 \kappa^2 \Lambda)^{3/2} \sqrt{1-4 \kappa^2 \Lambda \pm \sqrt{1-16 \kappa^2 \Lambda}}}.~~
\end{eqnarray}
\par Much of our understanding of nonlinear systems derives from the linearization about critical points
and from the determination of existing invariant submanifolds, which are structures that organize
the dynamics in phase space. The system under examination here presents a two-dimensional invariant
manifold of the dynamics defined by
\begin{eqnarray}
\label{eqInv}
p_M=p_N=p_R,~~~M=N=R.
\end{eqnarray}
This invariant plane is actually the intersection of two four-dimensional invariant submanifolds, defined by
$(M=N,~ p_M=p_N)$ and $(N=R,~p_N=p_R)$. The critical points obviously belong to the invariant plane.
\par Finally a straightforward analysis of the infinity of the phase space
shows that it has two critical points in this region, one acting as an attractor (stable  de Sitter configuration)
and the other as a repeller (unstable de Sitter configuration).
The scale factors $M$, $N$ and $R$ approach the de Sitter attractor as $M=N=R \sim \exp (\sqrt{\Lambda/3}~t)$,
so that the two de Sitter configurations also belong to the invariant plane. The phase picture of the invariant
plane is displayed in Fig. \ref{InvPlane}, in the variables $(x,p_x)$ defined in section \ref{sectionV}.
\par To proceed let us now linearize the dynamical equations (\ref{eqH}) about the critical
points $(M=N=R=M_{0_{i}},~p_M=p_N=p_R=0)$, $i=1,~2$. Defining
\begin{eqnarray}
\label{eqLine1}
\nonumber
X&=&(M-M_{0_{i}}),~~~~W=(p_M-0),\\
Y&=&(N-M_{0_{i}}),~~~~K=(p_N-0),\\
\nonumber
Z&=&(R-M_{0_{i}}),~~~~L=(p_R-0),
\end{eqnarray}
we obtain
\begin{eqnarray}
\label{eq9}
\left(
\begin{array}{c}
\dot{X}  \\
\dot{Y}   \\
\dot{Z}   \\
\dot{W}   \\
\dot{K}   \\
\dot{L}
\end{array}
\right)=\left(
\begin{array}{cccccc}
0 & 0 & 0 & -\alpha & \alpha & \alpha  \\
0 & 0 & 0 & \alpha & -\alpha & \alpha  \\
0 & 0 & 0 & \alpha & \alpha & -\alpha  \\
\beta & \gamma & \gamma & 0 & 0 & 0 \\
\gamma & \beta & \gamma & 0 & 0 & 0 \\
\gamma & \gamma & \beta & 0 & 0 & 0
\end{array}
\right)
\left(
\begin{array}{c}
X  \\
Y   \\
Z   \\
W   \\
K   \\
L
\end{array}
\right)
\end{eqnarray}
where
\begin{eqnarray}
\label{eqLine2}
\nonumber
\alpha=\frac{1}{4M_{0_i}},~~\beta=\frac{3}{M_{0_i}}-\frac{2\kappa^2E^2_{{\rm cr}_{(i)}}}{M^5_{0_i}},\\
\gamma=2\Lambda M_{0i}-\frac{3}{2M_{0_i}}-\frac{\kappa^2E^2_{{\rm cr}_{(i)}}}{M^5_{0_i}}.
\end{eqnarray}
The associated characteristic polynomial results
\begin{eqnarray}
\label{eqLine3}
\nonumber
P(\lambda)=(\lambda-\sqrt{2\gamma\alpha+\beta\alpha}~)(\lambda+\sqrt{2\gamma\alpha+\beta\alpha}~)\\
\times (\lambda-\sqrt{2\alpha(\gamma-\beta)}~)^2 ~(\lambda+\sqrt{2\alpha(\gamma-\beta)}~)^2,
\end{eqnarray}
with roots
\begin{eqnarray}
\label{eqLine5}
\lambda_{(i)}&=&\pm i\frac{\sqrt{2}}{M_{0i}}~,\\
\lambda_{(i)}&=&\pm \sqrt{3\Lambda-\frac{1}{2M^2_{0i}}}~,
\label{eqLine4}
\end{eqnarray}
where (\ref{eqCr}) was used. The pair of imaginary eigenvalues (\ref{eqLine5}) has multiplicity two,
characterizing a center-center structure. The analysis of the center-center structure will
reveal a manifold of linearized unstable periodic orbits with the topology of $S^3$. The extension
of this manifold to the nonlinear domain constitutes the center manifold\cite{wiggins01,guckenheimer}
of unstable periodic orbits,
parametrized with the constant of motion $E_0$ (with $E_0 < E_{\rm cr}$), which will play a central
role in our discussions in the next section.
\par Using (\ref{eqHC}) and (\ref{eqM012}) one can show that the pair of eigenvalues (\ref{eqLine4})
are imaginary for the critical point $i=1$ and real for the critical point $i=2$.
As we shall see below, we have (in the latter case $i=2$) that the critical
point $P_2$ is a saddle-center-center about which the dynamics has the
topology $R \times S^3$. On the other hand, the critical point $P_1$ is a center-center-center critical point,
about which the dynamics has the topology $S^1 \times S^3$ corresponding to perpetually oscillatory Bianchi IX
universes.
\par
Finally we should note that, in the limit case of a single critical
point (when $16 \kappa^2 \Lambda=1$), the second pair of eigenvalues (\ref{eqLine4}) are zero and no saddle
structure is present in the dynamics.
The analysis of this case will not be undertaken here. In the remaining of
this section our discussion follows the lines of \cite{ozorio} done for the
general relativity case.
\par To display the structure of the linearized motion, we start by diagonalizing the linearization matrix of (\ref{eq9})
with the use of a similarity transformation $\Re$ whose columns are composed of six independent eigenvectors of
the linearization \cite{siegel}. A judicious choice of $\Re$ yields primed variables defined by the transformation
\begin{eqnarray}
\label{eqJudicious}
\nonumber
X^{\prime}&=&\frac{1}{3}~ (X+Y+Z),\\
\nonumber
Y^{\prime}&=&\frac{1}{M_{0_i}}~ (X-Y),\\
Z^{\prime}&=&\frac{1}{M_{0_i}}~ (X+Y-2 Z),\\
\nonumber
W^{\prime}&=& (W+K+L),\\
\nonumber
K^{\prime}&=&\frac{M_{0_i}}{2}~(W-K),\\
\nonumber\
L^{\prime}&=&\frac{M_{0_i}}{6} ~(W+K-2 L).
\end{eqnarray}
In these new variables, the quadratic Hamiltonian about the $i$-th critical point
is expressed in the form
\begin{eqnarray}
\label{eqLinexyz}
\nonumber
H&=&\frac{1}{4}\Big(\frac{1}{6}~{W^{\prime}}^2-6 q {X^{\prime}}^2 \Big)
-\Big(\frac{1}{2 M_{0_i}^3} {K^{\prime}}^2+ M_{0_i} {Y^{\prime}}^2 \Big)\\
&-&\Big(\frac{3}{2 M_{0_i}^3} {L^{\prime}}^2+ \frac{M_{0_i}}{3} {Z^{\prime}}^2 \Big)+2(E_{\rm cr_{(i)}} -  E_0),
\end{eqnarray}
where
\begin{eqnarray}
\label{eq_q}
q=6 (\Lambda  M_{0_i}- \kappa^2 E_{\rm cr_{(i)}}^2/ M_{0_i}^5).
\end{eqnarray}
\par These primed variables are conjugated to the pairs according to $[X^{\prime},W^{\prime}]=1,~ [Y^{\prime},K^{\prime}]=1,~ [Z^{\prime},L^{\prime}]=1$,
other Poisson brackets (PB) zero. The Hamiltonian (\ref{eqLinexyz}) is separable, and we can identify the following
constants of the linearized motion
\begin{eqnarray}
\label{eqConst}
E_{q}&=&\frac{1}{4}\Big(\frac{1}{6}~{W^{\prime}}^2-6 q {X^{\prime}}^2 \Big),\\
E_{rot1}&=&\frac{1}{2 M_{0_i}^3} {K^{\prime}}^2+ M_{0_i} {Y^{\prime}}^2 ,\\
E_{rot2}&=&\frac{3}{2 M_{0_i}^3} {L^{\prime}}^2+ \frac{M_{0_i}}{3} {Z^{\prime}}^2,\\
Q_1&=&\frac{M_{0_i}}{3} Y^{\prime}Z^{\prime}+\frac{1}{2M_{0_i}^3} K^{\prime}L^{\prime},\\
Q_2&=&\frac{1}{2 M_{0_i}}\Big( L^{\prime}Y^{\prime}-\frac{1}{3}K^{\prime}Z^{\prime}\Big),
\end{eqnarray}
in the sense that they all have zero PB (\ref{eqLinexyz}). The first three
constants appear as separable pieces in the Hamiltonian (\ref{eqLinexyz}).
\par
The case of $E_q$ demands a separate analysis for the two critical points. From
previous relations we have that $q >0$ for the critical point $P_2$, so that
$E_{q}$ corresponds to the energy associated with the motion in the saddle sector.
We remind that this is connected to the fact that the second pair of eigenvalues (\ref{eqLine4})
are real for $P_2$. For the critical point $P_1$ in which $q<0$, $E_q$ corresponds to the rotational
energy in the additional rotational sector of the dynamics about $P_1$ which has the
structure $S^1 \times S^3$ as mentioned already.
The center-center-center critical point $P_1$ corresponds to a stable Einstein universe configuration
with perpetually oscillatory orbits in its neighborhood.
\par
In the following our focus will be the dynamical phenomena connected to the presence of the saddle-center-center critical point $P_2$
in the phase space of the model. We remark however that the analysis of the center manifold of unstable periodic
orbits can also be applied to the case of the critical point $P_1$.

\section{The dynamics about the saddle-center-center critical point\label{sectionV}}

We will now proceed to describe the topology of the general dynamics in the linear neighborhood of the
saddle-center-center critical point $P_2$ for which $q>0$.
\par
If $E_q=0$ two possibilities arise.
The first possibility is $W^{\prime}=0=X^{\prime}$. The total energy in this case is
$E_{rot1}+E_{rot2}$, the sum of the energies of the rotational
motion in the linear neighborhood of the center-center
manifold, corresponding to the motion on 2-dim tori\cite{berry}. The remaining
two constants $Q_1$ and $Q_2$ are additional symmetries that arise due to the multiplicity two of the
imaginary eigenvalues and are connected to the fact that the linearized dynamics in the center-center
sector is that of a 2-dim isotropic harmonic oscillator. They are not all independent but related by
\begin{eqnarray}
\label{eqRel}
4 E_{rot1} E_{rot2}= 12 Q_1^2+6 Q_2^2.
\end{eqnarray}
The motion in the constant energy surfaces $W^{\prime}=0=X^{\prime}$ are periodic orbits of the
2-dim isotropic harmonic oscillator, with Hamiltonian
\begin{eqnarray}
\label{eqHH}
\nonumber
H&=&\Big(\frac{1}{2 M_{0_i}^3} {K^{\prime}}^2+ M_{0_i} {Y^{\prime}}^2 \Big)+\Big(\frac{3}{2 M_{0_i}^3} {L^{\prime}}^2
+\frac{M_{0_i}}{3} {Z^{\prime}}^2 \Big)\\
&-&2(E_{\rm cr_{(i)}}-E_0)=0.
\end{eqnarray}
The above equation shows that $E_0 -E_{cr} <0 $ is necessary for the dynamics in the
rotational sector, defining a condition for the existence of the center-center manifold of periodic orbits.
\par By a proper canonical rescaling of the variables in (\ref{eqHH}) we can see that
these constant energy surfaces are hyperspheres and that the constants of motion
$Q_1$, $Q_2$ and $Q_3=E_{rot_1}-E_{rot_2}$ satisfy the algebra of the 3-dim rotation group
under the PB operation, namely,
\begin{eqnarray}
\label{eqAlg}
[Q_i,Q_j]=\epsilon^{ijk}Q_k.
\end{eqnarray}
The constant of motion $Q_1$ considered as a generator of infinitesimal contact transformations
has a peculiar significance in characterizing the topology of the underlying group of the
algebra (\ref{eqAlg}). While $Q_2$ generates infinitesimal rotation of the orbits,
$Q_1$ generates infinitesimal changes in eccentricity. The action of $Q_1$ is to take an orbit --
let us say nearly circular -- and to transform it into an orbit of higher and higher eccentricity
until it collapses into a straight line. Continued application of $Q_1$ produces again an elliptic orbit,
but now traversed in the opposite sense, so that it takes a $720^{\circ}$ to bring the orbit back
into itself. The two-valuedness  of the mapping arises from the fact that the orbits are oriented.
Therefore the group generated by these constants of motion is homomorphic to the unitary unimodular
group\cite{mcintosh} so that the topology of the center-center manifold is in fact $S^3$.
\par Due to the separate conservation of $E_{rot_1}$ and $E_{rot_2}$ (cf. (\ref{eqHH})) one can show
that the center manifold in the linear neighborhood of the critical points is foliated by
Clifford 2-dim surfaces in $S^3$\cite{sommer}, namely, 2-tori $\Im_{E_0}$ contained in the energy
surface $E_{0}={\rm const}.$ Such surfaces, as well as the $S^3$ manifold containing them, depend
continuously on the parameter $E_0$. We remark that these two tori will have limiting configurations
$E_{rot_1}=0$ or $E_{rot_2}=0$, and correspond to the case of maximum eccentricity (for instance, a straight line
in the plane $(Y^{\prime},Z^{\prime}))$.
\par The second possibility to be considered is $W^{\prime}= \pm 6 \sqrt{q} X^{\prime}$.
It defines the linear stable $V_S$ and unstable
$V_U$ manifolds of the saddle sector. $V_S$ and $V_U$ limit regions $I$ $(E_q \equiv E_{hyp}<0)$ and
regions $II$ $(E_q \equiv E_{hyp} > 0)$ of motion on hyperbolae which are solutions in the separable saddle sector
$E_{hyp}=\frac{1}{4}(\frac{1}{6}~{W^{\prime}}^2-6 q {X^{\prime}}^2 )$. Note that the saddle sector depicts the
structure of the neighborhood of $P_2$ in Fig. 1, with $V_U$ and $V_S$ tangent to the separatrices at $P_2$.
The direct product of $\Im_{E_0}$ with $V_S$ and $V_U$ generates, in the linear neighborhood of the critical point
($i=2$) the structure of stable ($\Im_{E_0} \times V_S$) and unstable ($\Im_{E_0} \times V_U$) 3-dim tubes
which coalesce, with an oscillatory approach to the tori $\Im_{E_0}$ for $t \rightarrow \infty$.
The energy of any orbit on these tubes is the same as that of the orbits on the tori
$\Im_{E_0}$. These structures are contained in the 4-dim energy surface $H=E_0$ such that $(E_0-E_{cr})<0$.
We should recall that the tubes constitute a boundary for the general flow and are defined by $E_q=0$ in the
linear neighborhood of the critical point. Depending on the sign of $E_q$ the motion will be confined inside
the 4-dim tube (for $E_q<0$) and will correspond to a flow separated from the one outside the tube (for $E_q >0$).
A detailed examination of the above motion and its extension to the nonlinear regime will be done in the next section.

\par
\begin{figure}
\begin{center}
\includegraphics[width=10cm,height=8cm]{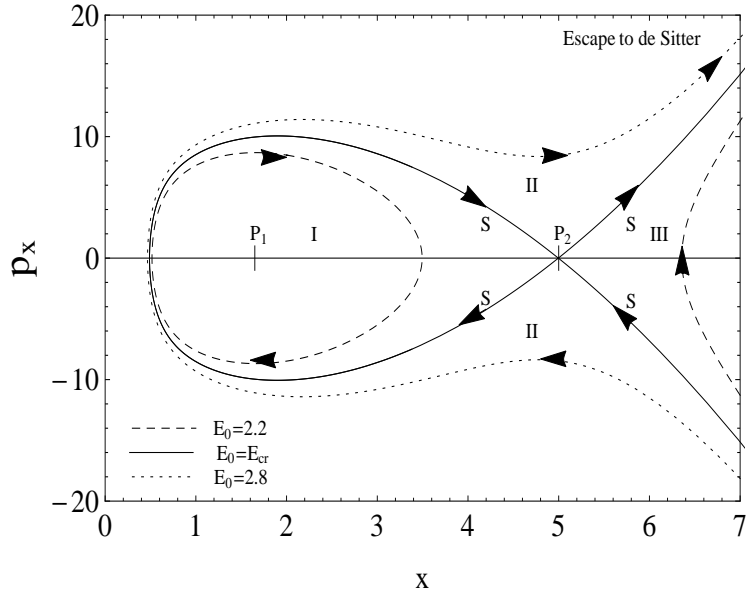}
\caption{The phase portrait of the invariant plane for $\kappa^2=0.5/E_0^2$ and $\Lambda=0.01$.
The critical points $P_1$ and $P_2$ belong to the invariant plane. The separatrices $S$ emerging
from $P_2$ separate the plane in three disconnected regions, one of oscillatory universes
the other two of one bounce universes. The separatrix dividing the regions $I$ and $II$
makes a homoclinic connection with $P_2$.
The linear neighborhood of $P_2$ depicts the motion of the saddle sector
with $V_S$ and $V_U$ tangent to the separatrix at $P_2$.
The figure is constructed in the canonical coordinates defined in (\ref{eqGen}).
}
\label{InvPlane}
\end{center}
\end{figure}

 The extension of our analysis beyond a linear neighborhood of critical points could be
made by implementing normal forms\cite{arnold,murdock} and associated coordinates, modulo their radius of convergence.
We will instead propose here a suitable canonical transformation which will
allows us to obtain an exact analytical form for the center manifold as well as
a sufficiently accurate description of the phase space dynamics
in extended regions away from the critical points. In particular we can examine the behavior of the nonlinear extensions
$W_S$ and $W_U$ of, respectively, the linear stable ($\Im_{E_0} \times V_S$) and linear unstable
($\Im_{E_0} \times V_U$) manifolds\cite{wiggins02} emanating from the neighborhood
of the saddle-center-center $P_2$.
Let us introduce the canonical transformation with the generating function
\begin{eqnarray}
\label{eqGen}
G=(M N R)^{1/3} p_{x}+\frac{M}{N} p_{y}+\frac{MN}{R^2} p_{z},
\end{eqnarray}
where $p_{x}$, $p_{y}$ and $p_{z}$ are the new momenta, resulting in
\begin{eqnarray}
\label{eqGen1}
x=(M N R)^{1/3},~~~y=\frac{M}{N},~~~z=\frac{MN}{R^2},
\end{eqnarray}
and
\begin{eqnarray}
\label{eqGen2}
\nonumber
p_M&=&\frac{1}{3}\frac{NR}{(MNR)^{2/3}}p_x+\frac{1}{N}p_y+\frac{N}{R^2}p_z,\\
p_N&=&\frac{1}{3}\frac{MR}{(MNR)^{2/3}}p_x-\frac{M}{N^2}p_y+\frac{M}{R^2}p_z,\\
\nonumber
p_R&=&\frac{1}{3}\frac{MN}{(MNR)^{2/3}}p_x-\frac{2MN}{R^3}p_z.
\end{eqnarray}
Here, the variable $x$ is obviously the average scale factor of the model.
In these new canonical variables the equations of the invariant plane reduce to
\begin{eqnarray}
\label{eqInv2}
y=1,~~~z=1,~~~p_y=0=p_z.
\end{eqnarray}
It is then clear that $(x,p_x)$ are variables defined on the invariant plane. In these variables
the phase space picture of the invariant plane is given in Fig. \ref{InvPlane}.
The separatrices $S$ emerging from the saddle-center-center $P_2$ separate the invariant plane
in three disconnected regions, region $I$ of oscillatory universes and regions $II$ and $III$
of one bounce universes. They are constituted of three branches, namely, the separatrix that divides
the regions $I$ and $II$ and makes a homoclinic connection with the critical point $P_2$ and two
others that approach the de Sitter asymptotic configurations for $t \rightarrow \pm \infty$.
The first branch will play a fundamental role in our following discussions and will be
referred to as separatrix, except where a qualification is needed to avoid confusion.
The center-center-center $P_1$ corresponds to a
stable Einstein universe configuration that occurs due to the bulk-brane interaction term
proportional to $E_0^2$ in the Hamiltonian (\ref{eq4}).
\par In the variables $(x, p_x, y, p_y, z, p_z)$ the full Hamiltonian (\ref{eq4}) assumes the form
\begin{eqnarray}
\label{eqGen21}
\nonumber
&&H(x, p_x, y, p_y, z, p_z;E_0)=\frac{1}{24 x}p_{x}^{2}-\frac{y^2}{2 x^3}p_{y}^{2}\\
\nonumber
&-&\frac{3z^2}{2x^3}p_{z}^{2}
-\frac{x}{2 z^{4/3}}-\frac{1}{2}x z^{2/3}y^{2}
-\frac{1}{2 y^2}x z^{2/3}+\frac{x}{y z^{1/3}}\\
&+&\frac{xy}{z^{1/3}}
+x z^{2/3} +\frac{\kappa^2E_0^2}{x^3}-2 \Lambda x^3-2E_0=0.
\end{eqnarray}
We remark that the linearization of (\ref{eqGen1})-(\ref{eqGen21}) about both
critical points $P_1$ and $P_2$ of the dynamical system (\ref{eqH}) yields exactly the transformation
(\ref{eqJudicious}), and that the variables $(y,p_y,z,p_z)$ correspond to the
primed variables $(K^{\prime},Y^{\prime},L^{\prime},Z^{\prime})$ defined on the
center-center manifold $S^3$ about a linear neighborhood of $P_2$.
\par The new canonical variables are most convenient since they separate the
degrees of freedom of the system into pure rotational modes, $(y,p_y)$ and $(z,p_z)$,
and the expansion/contraction mode $(x,p_x)$ connected to the invariant plane.
This can be illustrated by implementing the expansion of the
dynamical system generated from (\ref{eqGen21}) about a linear neighborhood of the invariant plane,
producing a linearized Hamiltonian parametrized by the variables $(x(t),p_x(t))$ describing the
curves in the invariant plane, for instance, in the region $I$ of periodic orbits bounded by the
separatrix $S$ homoclinic to $P_2$. This is analogous to the usual expansion
of a dynamical system about a periodic orbit. Using (\ref{eqGen21}), we then obtain
\begin{eqnarray}
\label{eqHparam}
\nonumber
H&=&{\mathcal E}_{\rm inv}-\frac{1}{2 x^3}(p_y^2+3 p_z^2)-x(y-1)^2\\
&-&\frac{1}{3}x(z-1)^2=2 E_0,
\end{eqnarray}
where
\begin{eqnarray}
\label{eqHparam1}
{\mathcal E}_{\rm inv}=\frac{1}{24 x}p_{x}^{2} +\frac{3x}{2}+\frac{\kappa^2 E_0^2}{x^3}-2 \Lambda x^3 = {\rm const.}
\end{eqnarray}
The resulting dynamical equations are
\begin{eqnarray}
\label{eqDSparam}
\nonumber
\dot{\delta y}&=&-\frac{1}{x^3}~ \delta p_y,\\
\nonumber
\dot{ p_y}&=&2 x~ \delta y,\\ \\
\nonumber
\dot{\delta z}&=&-\frac{3}{x^3}~ \delta p_z,\\
\nonumber
\dot{\delta p_z}&=&\frac{2 x}{3}~ \delta p_y,
\end{eqnarray}\\
where $\delta y=(y-1)$, $\delta z=(z-1)$, $\delta p_y=(p_y-0)$, and $\delta p_z=(p_z-0)$. The linearization matrix of (\ref{eqDSparam})
has imaginary eigenvalues $\lambda= \pm i \sqrt{2}/x(t)$, both with multiplicity two, corresponding to
elliptical modes in the linear neighborhood of the invariant plane so that the motion is oscillatory about the
invariant plane.

\section{The non linear center manifold and the homoclinic cylinders\label{sectionVI}}

\par The nonlinear extension of the center manifold, by continuity, maintains the topology
$S^3$ but it can no longer be decomposable into $E_{rot_1}$ and $E_{rot_2}$ so that now
only the 4-dim tubes with the topology $R \times S^3$ are meaningful for the nonlinear dynamics.
Similarly the extension of the structure of the 4-dim tubes away from the neighborhood of the
center manifold are to be examined, and our basic interest will reside in the stable and unstable
tubes, $W_S=V_S \times S^3$ and $W_U=V_U \times S^3$, that leave this neighborhood. The tubes have
the structure of 4-dim spherical cylinders (of co-dimension $2$), one less dimension than the energy surface, and act
therefore as separatrices, separating the energy surface in two dynamically disconnected parts. The 2-dim invariant plane, defined
by (\ref{eqInv}), is contained in a 6-dim phase space and it is obvious that, contrary to examples
in lower dimensional systems, it does not separate the dynamics in disjoint parts. In fact the
general motion about the curves of the invariant plane is an oscillatory flow confined in the interior
or exterior of 4-dim tubes $R \times S^3$, so that the invariant plane (or more properly, one of
the curves of the invariant plane) can be thought as a structure in the center of the tubes. This latter fact
is of crucial importance in the discussion of the transversal crossing of the 4-dim cylinders $W_S$ and $W_U$
made in section \ref{sectionVII}.
\par The nonlinear extension of the center manifold in the canonical variables $(y,z,p_y,p_z)$
is obtained by substituting $x=x_{cr}$ and $p_x=0$ in (\ref{eqGen21}), yielding after some
manipulation the exact analytical expression
\begin{eqnarray}
\label{eqCenter}
\nonumber
H_c&=&\frac{y^2}{2 x_{cr}^3}p_{y}^2+\frac{3 z^2}{2 x_{cr}^3}p_z^2
+x_{cr}\Big(\frac{3}{2}+\frac{1}{2 z^{4/3}} \\
\nonumber
&+&\frac{1}{2} z^{2/3}y^{2}
+\frac{z^{2/3}}{2 y^2}-\frac{1}{y z^{1/3}}-\frac{y}{z^{1/3}}-z^{2/3} \Big)\\
&-& \kappa^2 \frac{E_0^2-E_{cr}^2}{x_{cr}^3}-2 (E_{cr}-E_0)=0,
\end{eqnarray}
\begin{figure}
\hspace{-2.0cm}
{\includegraphics*[height=6.5cm,width=8cm]{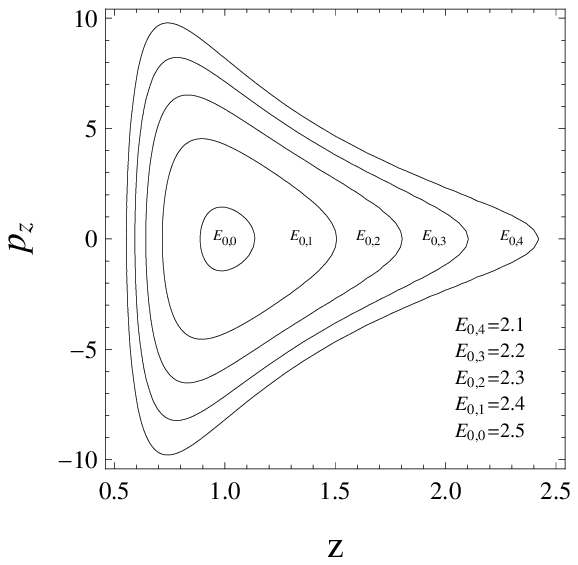}}~~~{\includegraphics*[height=6.5cm,width=8cm]{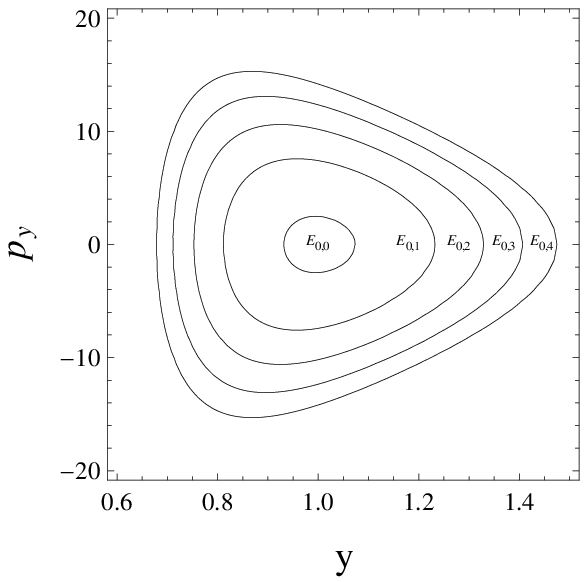}}
\caption{The sections $(y=1, p_y=10^{-3})$ (left panel) and $(z=1, p_z=0)$ (right panel) of the center manifold for five values
of $E_0$. We note the deformation of the center manifold $S^3$ in the nonlinear domain as $(E_{cr}-E_{0,i})$ increases.
Here $E_{cr}=2.5127254138199464$.}
\label{fig-zpz}
\end{figure}
where $x_{cr}$ and $E_{cr}$ are respectively the coordinate and the energy of the critical point
$P_2$.
The form (\ref{eqCenter}) adopted above for the center manifold equation makes explicit its
dependence on the parameter $(E_{cr}-E_0)$. For $E_0=E_{cr}$ the center manifold reduces to the
critical point. The domain of $E_0$ defining the center manifold
satisfies the constraint to $E_0 < E_{cr}$, as already discussed; the case of the linear version (\ref{eqHH})
corresponds to $(E_{cr}-E_0)$ sufficiently small. As $(E_{cr}-E_0)$ increases we have a nonlinear center manifold
parametrized by the energy $E_0$. In general the center manifold is a 3-dim submanifold of the 6-dim phase space
contained in the 5-dim energy hypersurface $H=E_0$.
In Figs. \ref{fig-zpz} we plot the sections $(y=1,p_y=10^{-3})$ and $(z=1,p_z=0)$ of the $S^3$ center manifold (\ref{eqCenter})
showing its deformation in the nonlinear regime as the values of $(E_{cr}-E_0)$ increase. We adopted the values $\Lambda=0.01$
and $\kappa^2=0.5$ so that the associated critical energy $E_{cr}=2.5127254138199464$ and $x_{cr}=4.974148895632555$
for the saddle-center-center $P_2$. In the Figures we selected five values for $E_0$.
\begin{figure}
\hspace{-1.5cm}
{\includegraphics*[height=5.5cm,width=5.5cm]{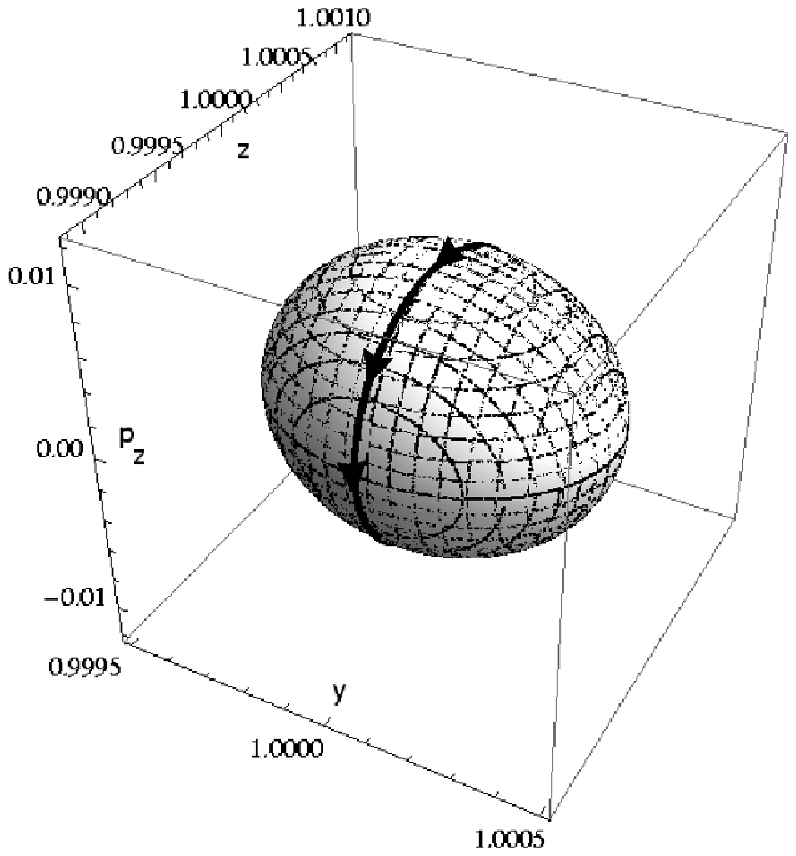}}{\includegraphics*[height=5.5cm,
width=5.5cm]{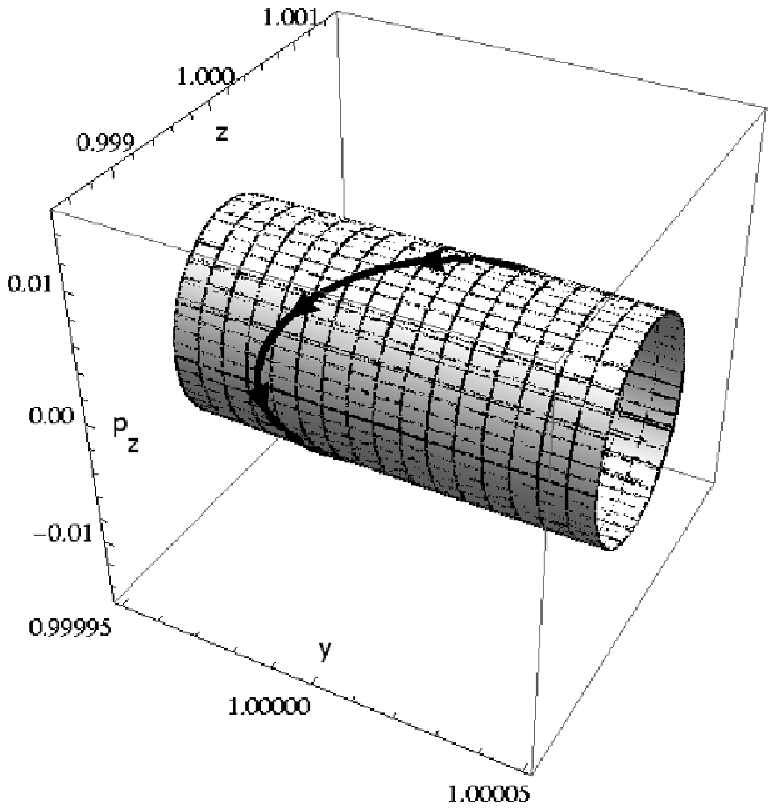}}{\includegraphics*[height=5.0cm,width=5.0cm]{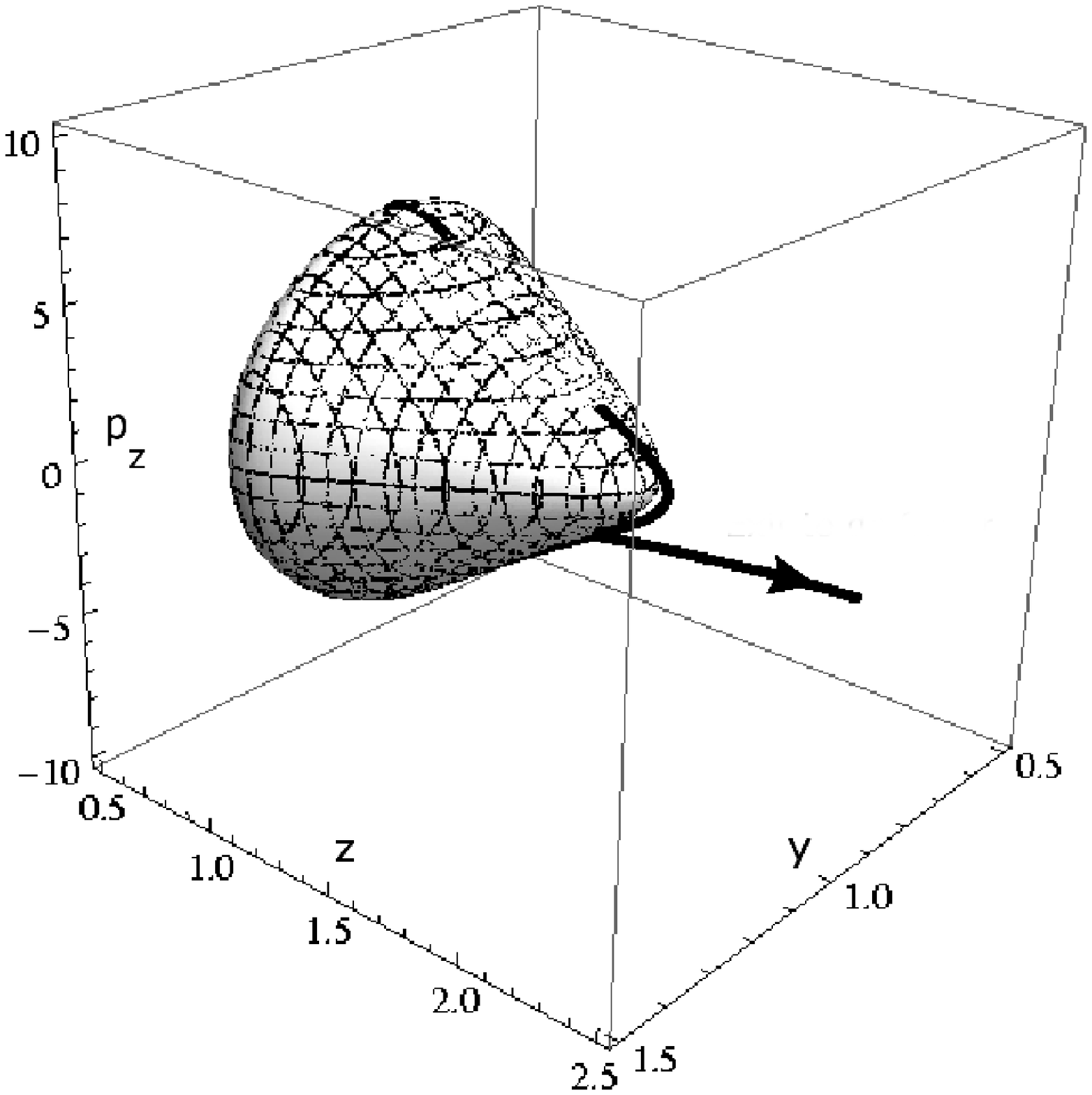}}
\caption{(left panel) The wireframe shows the 2-dim section $p_y=10^{-3}$ of the center manifold in the space $(y,z,p_z)$ for $E_0=2.512725$.
The solid line indicates an orbit for $0\leq t\lesssim 2000$ with initial conditions
$(x0 = xcr,y0 =1,z0 =1,px0 =0,py0 = 10^{-3},pz0 = 0.008177970306403768)$ on the center manifold.
(middle panel) An expanded piece of the center manifold on the left, including the same orbit.
(right panel) The wireframe shows the 2-dim section $p_y=10^{-3}$ of the center manifold in the nonlinear domain,
with $E_0=2.3$. The solid line corresponds to a one-bounce orbit which moves towards large values of $z$ (when $t \sim 37$, $z(t) \sim 63$) before
escaping to the de Sitter attractor when $t\simeq 70$. The initial condition for this orbit, ($x_0=x_{cr}$, $y_0=1$, $z_0=1$, $p_{x_{0}}=0$, $p_{y_{0}}=10^{-3}$, $p_{z_0}=5.879276478889785$) is taken on center manifold.
This orbit remains on the center manifold for a time up to $t \sim 12$. Here $E_{cr}=2.5127254138199464$.
}
\label{fig-3}
\end{figure}
\begin{figure}
\hspace{-1.8cm}
{\includegraphics*[height=5.0cm,width=5.5cm]{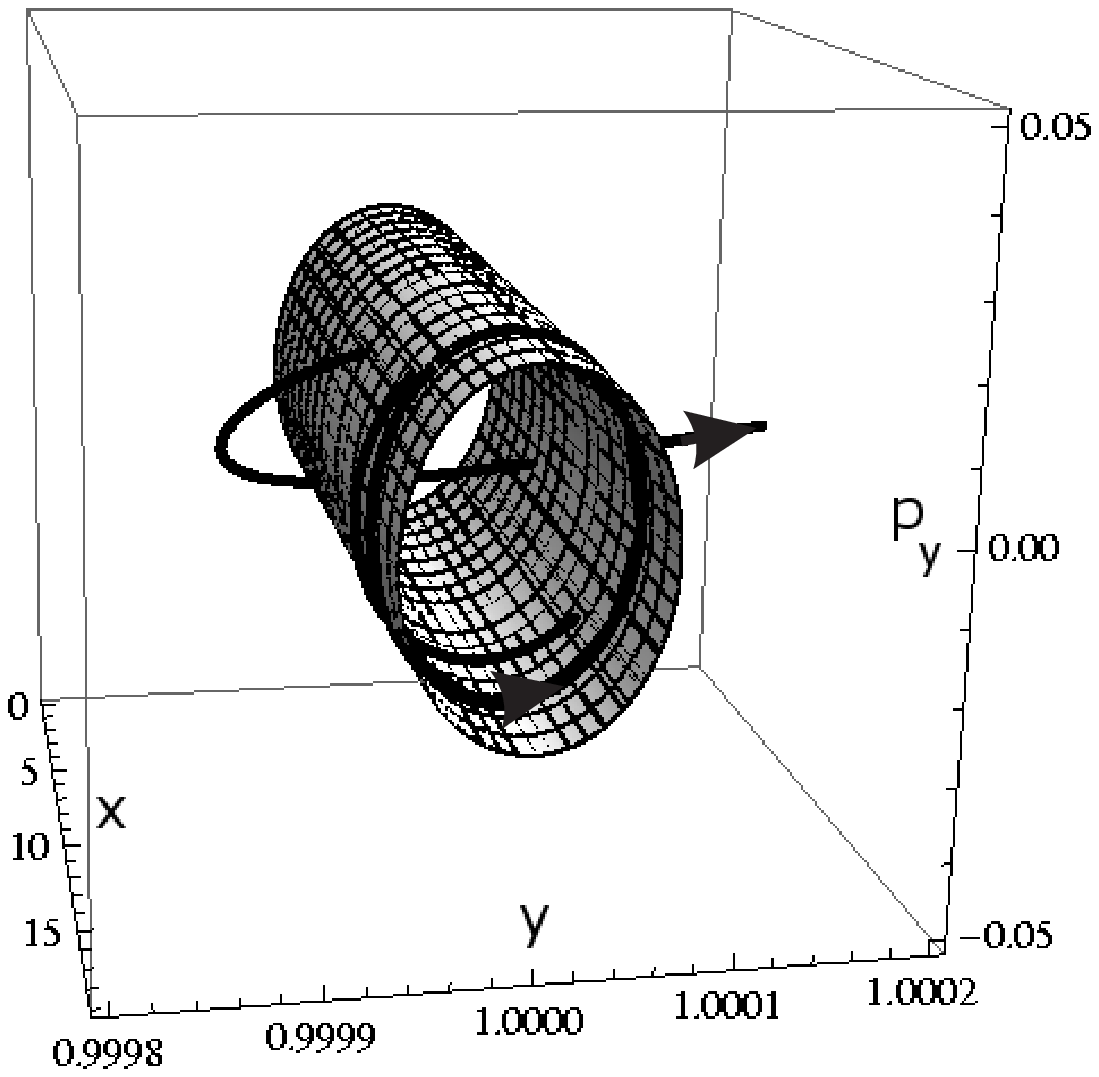}}~~~~~~~~{\includegraphics*[height=5.0cm,width=5.5cm]
{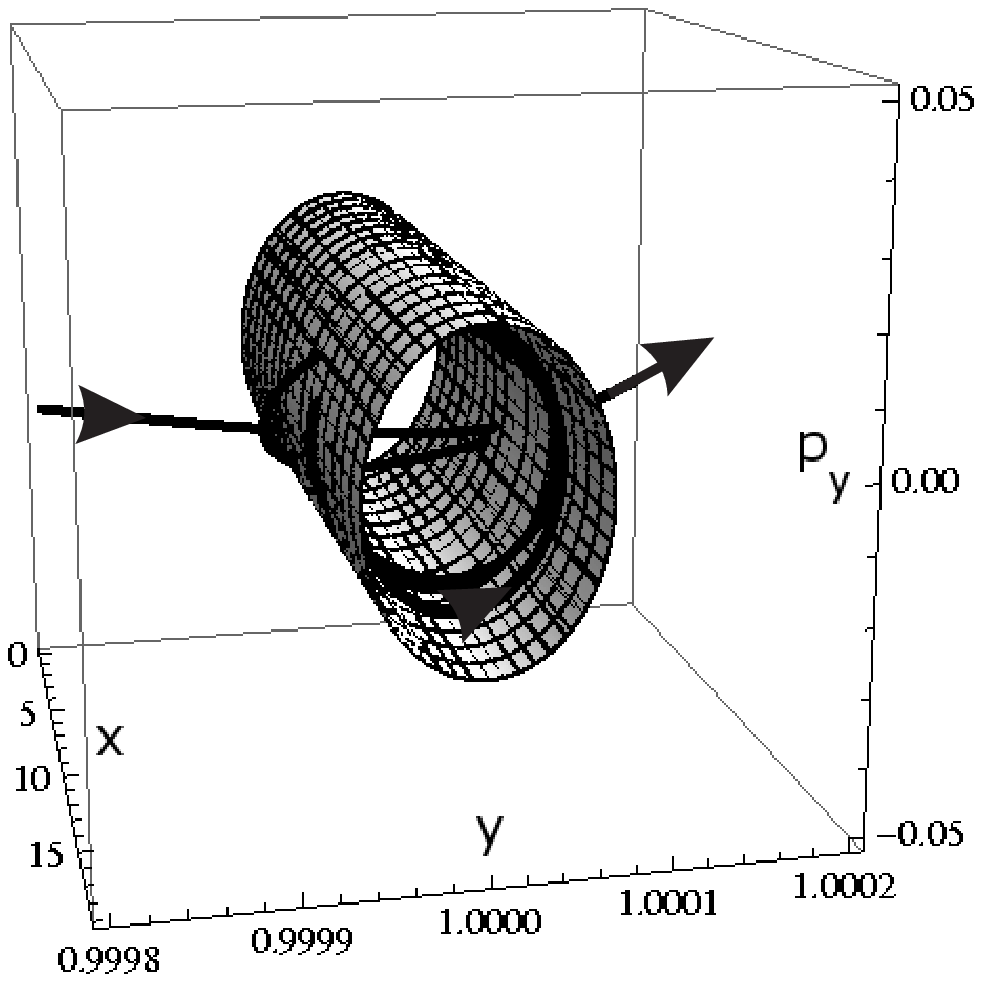}}~~~~{\includegraphics*[height=5.0cm,width=5.0cm]{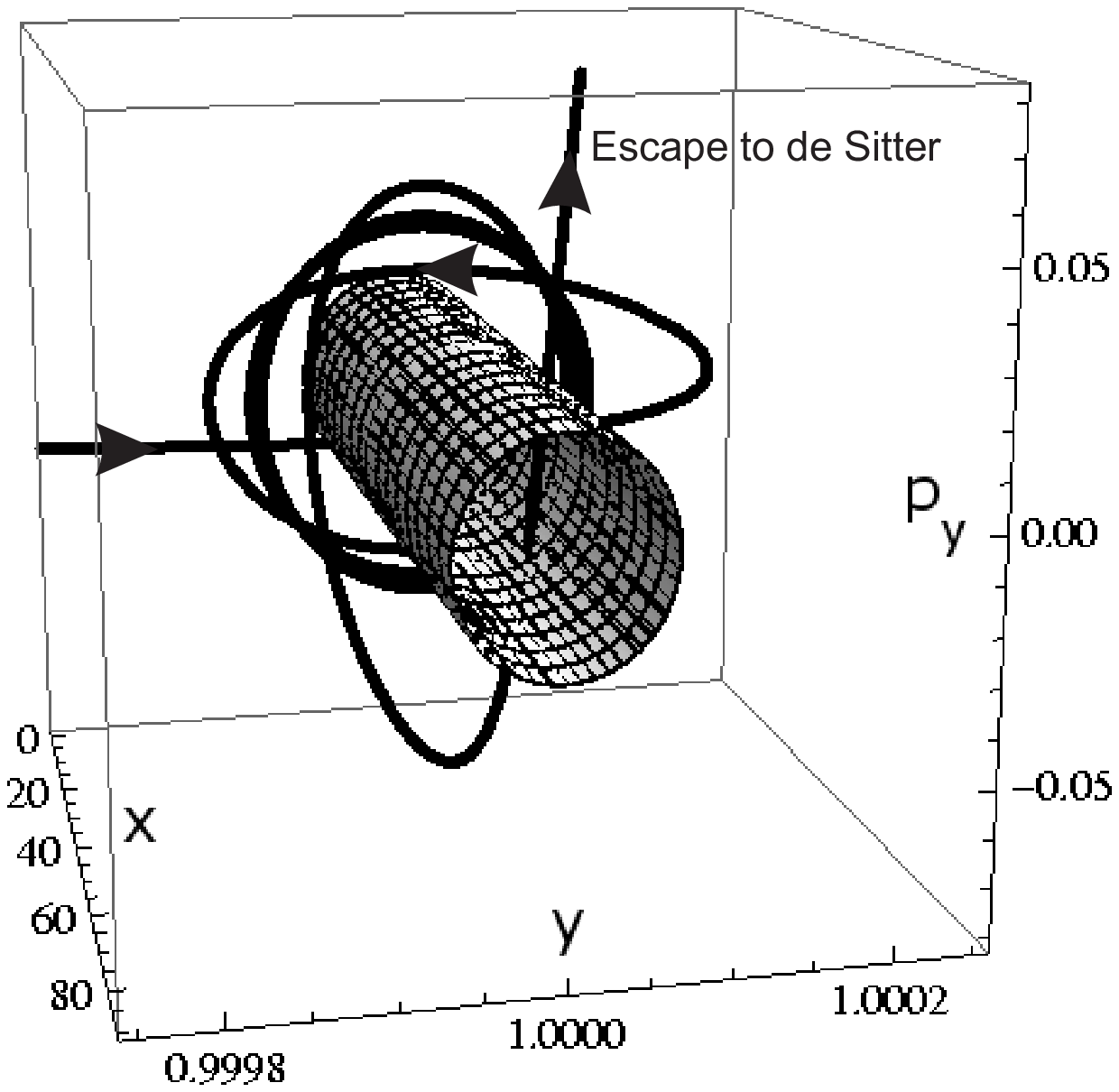}}
\caption{(left panel) The wireframe shows a piece of the 2-dim section $p_z=0$ of the center manifold in the coordinates $(x,y,p_y)$
for $E_0=7.9096539$. Here we fixed the parameters $\Lambda=0.001$ and $\kappa^2=0.5$, with corresponding $E_{crit}=7.909653935312942$.
The solid line is an orbit of the full dynamics for $0\leq t\leq 645$ with initial condition ($x_0=x_{cr_{(2)}}$, $y_0=1$, $z_0=1$, $p_{x_{0}}=0$, $p_{y_{0}}=0.023600434934712197$, $p_{z_0}=0$) taken on the center manifold.
The orbit remains on the center manifold up to $t \sim 500$ when it leaves its neighborhood and is driven towards its first bounce. (middle panel) The same
orbit of the left figure for $652 \leq t \leq 1320$. We can see that the orbit returns from its first bounce to a sufficiently small
neighborhood of the center manifold for $t\simeq 652$, before escaping to its second bounce when $t\simeq 1320$. (right panel) The solid line indicates
the same orbit of the previous figures but for $1330\leq t\leq 2020$. This orbit returns from its second bounce to a sufficiently
small neighborhood of the center manifold for $t\simeq 1330$ before escaping to the de Sitter attractor when $t\simeq 2020$.}
\label{fig-5}
\end{figure}
\begin{figure}
\hspace{-1.5cm}
{\includegraphics*[height=5cm,width=7.3cm]{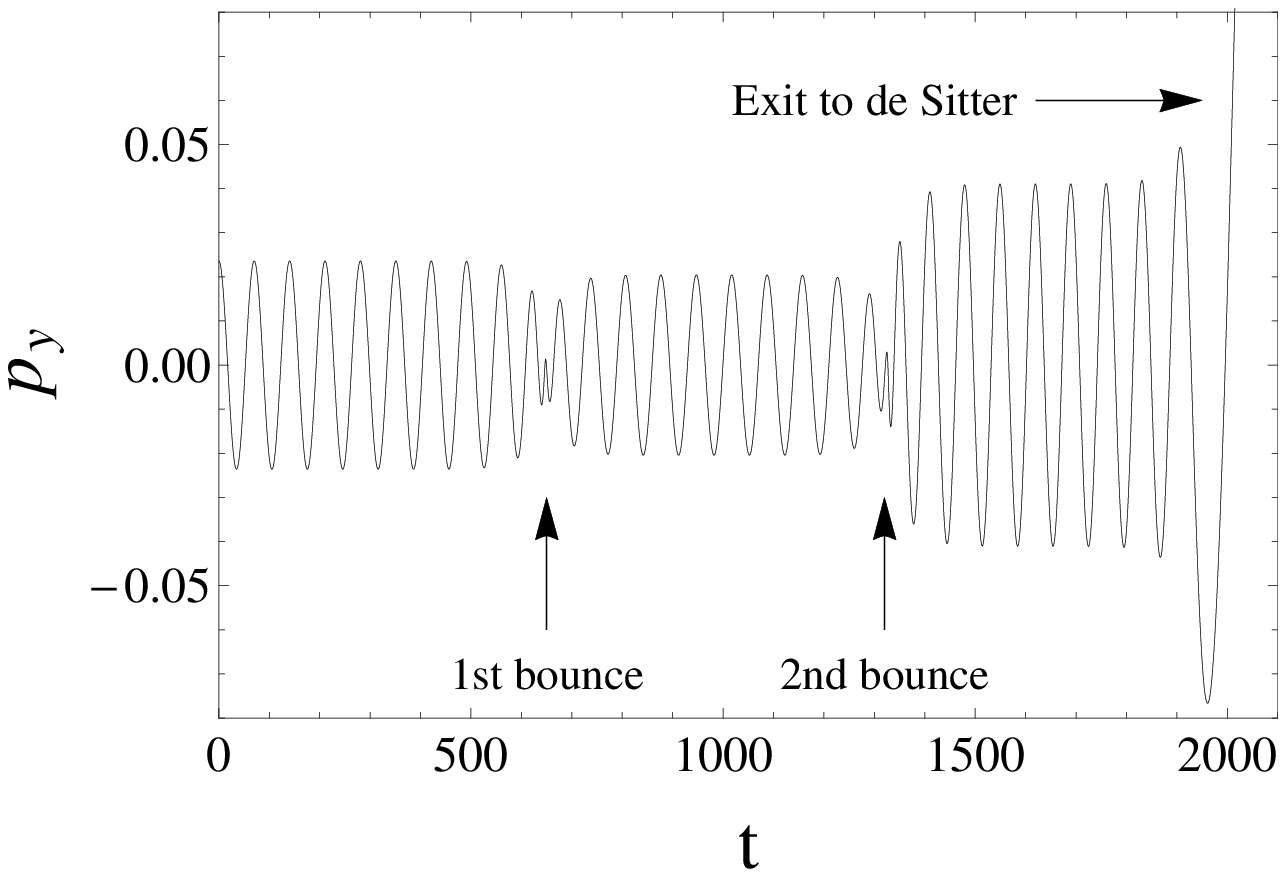}}~~~~{\includegraphics*[height=5cm,width=7.3cm]{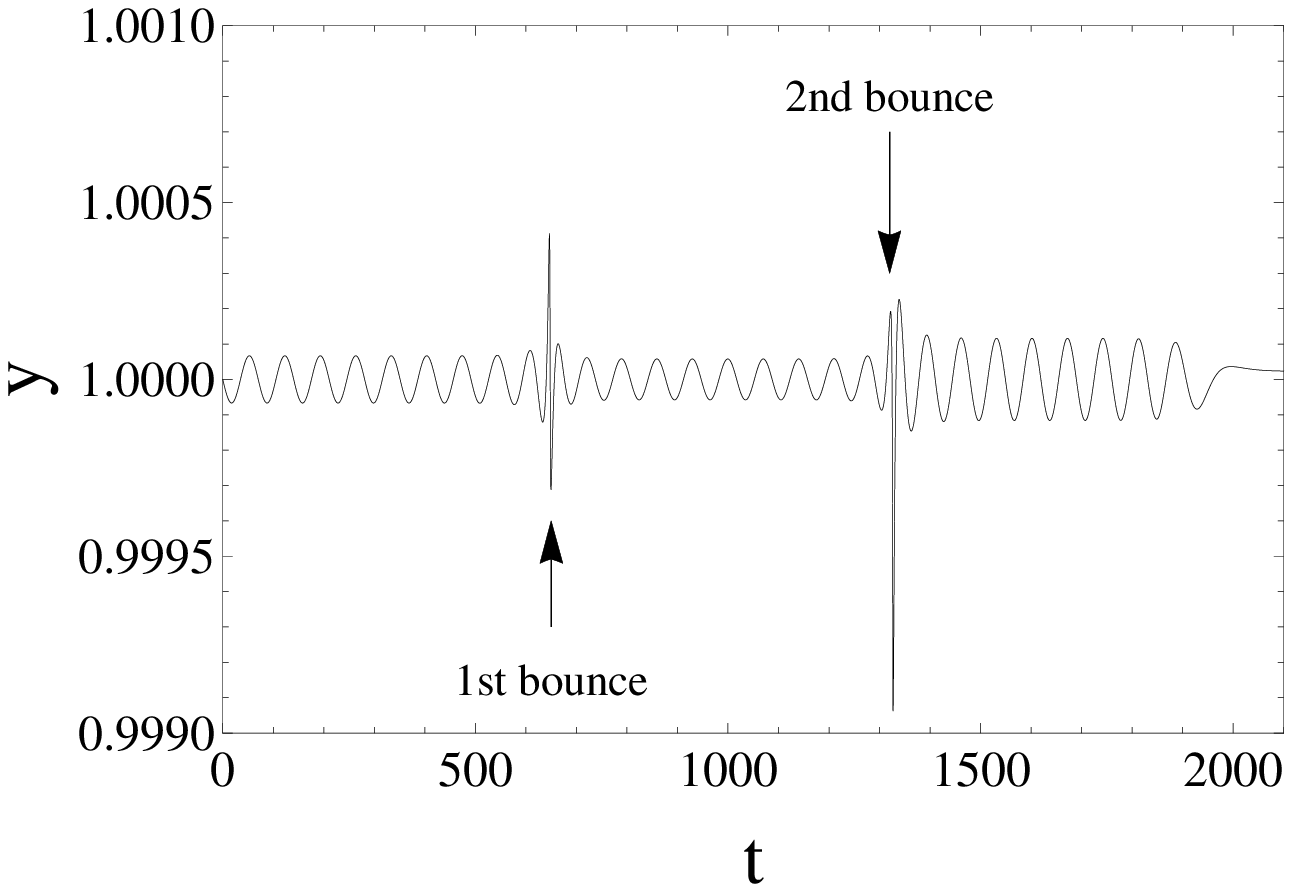}}
\caption{Plot of $p_y(t)$ and $y(t)$ corresponding to the orbit of Figs. \ref{fig-5} showing the typical oscillatory
behavior of the rotational modes in the dynamics of the model. We note the
decrease of the amplitude of $p_y(t)$ and the amplification of the conjugated $y(t)$
in a neighborhood of the bounce. This pattern is analogous
for the other variables of the orbit $(p_z,z)$.}
\label{fig-oscill}
\end{figure}
\par As we have already seen the canonical coordinates $(y, p_y, z, p_z)$ cover the center manifold $S^3$
and therefore we will use them not only to examine the stability of the motion restricted to $S^3$
but also to obtain an accurate description of whole the dynamics $(x(t), p_x(t), y(t), p_y(t), z(t), p_z(t))$
emerging from a neighborhood of the center manifold. In the following we will numerically illustrate this behavior.
We must remark that we do not make use here of the displacing (in the direction of the unstable cylinder)
of initial conditions taken on the invariant center manifold, as the shooting method in \cite{wiggins2},
but instead we make use of the instability of the motion on the center manifold which computationally conserves
the Hamiltonian constraint (\ref{eqGen21}) for all $t$.
Actually in all our numerical simulations the error in the Hamiltonian constraint (\ref{eqGen21})
is checked to remain $ \lesssim 10^{-13}$ for the whole computational domain.
\par To start let us fix the parameters $\kappa^2=0.5~,~\Lambda=0.01$ as in Figs. \ref{fig-zpz}. In Figs. \ref{fig-3}
we now show the 2-dim sections $p_y=10^{-3}$ of the center manifold for $E_0=2.512725$ (left).
The solid line indicates an orbit with initial conditions
obviously satisfying (\ref{eqCenter}).
This orbit is evolved with the full dynamics generated from the Hamiltonian (\ref{eqGen21})
and remains on the center manifold for $0 \leq t \lesssim 2000$.
A piece of this center manifold is displayed in Fig. \ref{fig-3} (middle) where the solid line describes the same previous orbit.
Fig. \ref{fig-3} (right) displays the section $p_y=10^{-3}$ of the center manifold for $E_0=2.3$.  The solid line corresponds to a one-bounce orbit
which moves towards large values of $z$ (when $t \sim 37$, $z(t) \sim 63$) before escaping to the de Sitter attractor when $t\simeq 70$.
The initial condition for this orbit
is taken on the center manifold $S^3$; it remains on the center manifold up to $t \sim 12$.
This increase of the dynamical instability is actually due to the large value of $(E_{crit}-E_0) \simeq 0.213$,
causing the orbit to leave rapidly the center manifold towards the bounce,
satisfying however the exact dynamics within an error $\lesssim 10^{-13}$.
\par A second set of experiments is displayed in Figs. \ref{fig-5} where we examined the oscillatory motion originating
in the 2-dim section $p_z=0$ of the center manifold
for $\Lambda=0.001$, $\kappa^2=0.5$ and $E_0=7.9096539$.
The continuous solid line shown in the left panel is an orbit of the full dynamics for $0\leq t\leq 645$
with the initial condition
taken on the center manifold. This oscillatory orbit, which is initially periodic, remains on the center manifold up to
$t \sim 500 $ when it leaves this neighborhood and is driven towards its first bounce. In the middle panel the same
orbit is shown for $652 \leq t \leq 1320$. We see that the orbit returns from the first bounce to a sufficiently small
neighborhood of the center manifold for $t\simeq 652$ before escaping to its
second bounce at $t\simeq 1320$. In the right panel we have the same orbit of the previous figures but for for $1330\leq t\leq 2020$.
This orbit returns from its second bounce to a sufficiently small neighborhood of the center manifold for $t\simeq 1330$, before
escaping to the de Sitter attractor when $t\simeq 2020$.
These numerical simulations also reveal a typical behavior of the dynamics as we decrease $E_0$. In fact
the increase of $(E_{crit}-E_0)$ makes an orbit, with initial conditions taken on the center manifold,
to rapidly leave this neighborhood indicating a dynamical instability (albeit the accuracy of the exact
dynamics) as shown in Fig. \ref{fig-3} (right).
\par We remark that the oscillatory behavior of the orbit in the phase space sectors $(y,p_y)$ and $(z,p_z)$
is typical, even when the orbit tends asymptotically to one of the deSitter attractors. This is illustrated in
Figs. \ref{fig-oscill} where we plot the time behavior of $p_y$ and $y$ of the orbit
discussed in Figs. \ref{fig-5}. We note a decrease of the amplitude of $p_y$
and an amplification of the amplitude for the conjugated $y$ in a neighborhood
of the bounce. this pattern is analogous for the other variables $(p_z,z)$ of the orbit.
\par
\begin{figure}
\hspace{3.3cm}
{\includegraphics*[height=6cm,width=6cm]{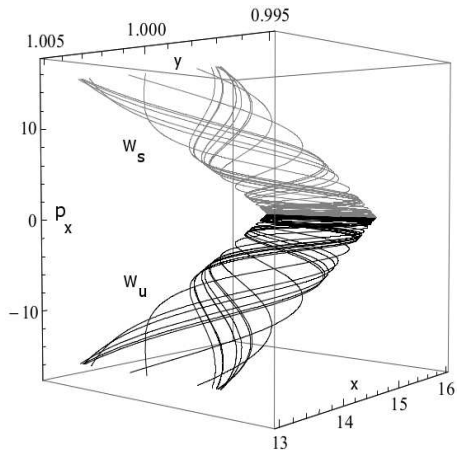}}
\caption{A numerical illustration of the unstable cylinder $W_U$ (black) spanned by $16$ orbits, with initial
conditions taken on a circle in the domain $(y,p_y)$ of the center manifold for $E_0=7.9096$ , emerging towards the bounce. From the same
domain of initial conditions the stable cylinder $W_S$ (gray) emerges towards the bounce. The projection of the figure
in the plane $(x,p_x)$ ``shadows'' the separatrix of the invariant plane.
Here $E_{crit}=7.909653935312942$.}
\label{fig-8}
\end{figure}

 Finally we give a numerical illustration of the stable and unstable cylinders
emanating from the center manifold which are a
nonlinear extension of the ${\Im}_{E_0} \times V_{S} $ and ${\Im}_{E_0} \times V_{U} $,
with ${\Im}_{E_0} \subset S^3$ defined in a linear neighborhood of the saddle-center-center $P_2$.
We must recall that these cylinders are actually composed
of orbits that have the same energy $(E_{cr}-E_0)$ of the center manifold and coalesce
to it as $t \rightarrow \pm \infty$. In Fig. \ref{fig-8} we display the stable $W_S$ and
unstable $W_U$ cylinders emanating from the neighborhood of the center manifold towards the bounce,
guided by the separatrix dividing the regions I and II of the invariant plane.
We emphasize that the separatix guiding the cylinders is actually a structure inside the cylinders.
We fixed the parameters $\Lambda=0.001$ and $\kappa^2=0.5$, as in Figs. \ref{fig-5}, and took $E_0=7.9096$
so that $(E_{{cr}}-E_0) \sim 10^{-5}$.
\par A comment is in order now. Since the cylinders $W_S$ and $W_U$ are 4-dim surfaces they obviously separate
the 5-dim energy surface defined by the Hamiltonian constraint (\ref{eqGen21}) in two dynamically disconnected pieces,
a fact that will be fundamental in the characterization of chaos in the case of an eventual transversal crossing
of $W_S$ and $W_S$\cite{wiggins02,guckenheimer}. We remark that in Fig. \ref{fig-8} the
projection of the cylinders on the plane $(x,p_x)$ ``shadows'' the separatrix
of the invariant plane, as expected since the separatrix is a structure contained in the interior
of the two tubes. As the separatrix in question makes an homoclinic connection to the saddle-center-center critical point
$P_2$ this fact leads necessarily to the transversal crossings of $W_S$ and $W_U$, a dynamical phenomenon that we
examine in the next section.

\section{The transversal crossings of the cylinders and the homoclinic intersection manifold: a chaotic saddle
 \label{sectionVII}}

The results of the previous sections showed that two 4-dim cylinders, one stable $W_S=R \times S^3$ and one unstable $W_U=R \times S^3$,
emerge from a neighborhood of the saddle-center-center $P_2$.
The center manifold $S^3$ is the locus of the rotational degrees of freedom of the phase space dynamics
and is parametrized with the energy $E_0$ ($E_0 < E_{crit}$. It encloses the critical point $P_2$
and tends to it as $E_0 \rightarrow E_{crit}$. At this limit the cylinders $W_S$ and $W_U$ reduce to the
separatrix $S$, which makes an homoclinic connection of $P_2$ to itself in the invariant plane.
The separatrix is a structure inside the cylinders, about which the flow with the
oscillatory degrees of freedom $(y,p_y,z,p_z)$ proceeds, guiding the cylinders towards the
bounce (cf. Fig.\ref{InvPlane}) and leading to their eventual crossing.
The first crossing is expected to occur in a neighborhood of the bounce $(x=x_b,p_x=0)$,
where $x_b$ is the scale factor of the bounce for the orbits at $p_x=0$.
In order to detect this first intersection we will adopt as the surface of section\cite{lieberman}
the 4-dim surface $\Sigma:(x=x_b,p_x=0)$.
This first transversal crossing of the cylinders will be the main object of the present section.
\par Due to the conservation of the Hamiltonian constraint (\ref{eqGen21}) we have that at the bounce
\begin{eqnarray}
\label{eqGen3}
H(x=x_b,p_x=0,y,p_y,z,p_z;E_0)=0,
\end{eqnarray}
which is the equation of a closed surface with the topology of $S^3$.
The transversal crossing of the stable $W_S$ and unstable $W_U$ cylinders at the bounce will therefore be
a set of points contained in the transversal intersection of two 3-spheres defined by (\ref{eqGen3}),
then a $S^2$. These points define a set of orbits that are contained both in
the stable and the unstable cylinders and are bi-asymptotic (homoclinic) to the center manifold $S^3(E_0)$.
They are denoted as homoclinic points and homoclinic orbits. Therefore the set of homoclinic points
has the compact support $S^2$. The presence of a homoclinic orbit in the dynamics
is an invariant signature of chaos in the model\cite{moser,wiggins01}.
The homoclinic intersection manifold has the topology $R \times S^2$
and consists of all homoclinic orbits biasymptotic
to the center manifold. In this sense, a {\it chaotic saddle}\cite{wiggins2} associated with $S^3$ is defined,
indicating that the set of intersection points of the cylinders has the nature of a Cantor-type set
with a compact support $S^2$\cite{cresson}.
\par A complete numerical study of the intersection of the 4-dim cylinders $W_S$ and $W_U$
is beyond the scope of the present paper (it will be considered as the subject for a future publication).
Here our numerical experiments will be restricted to the dynamics on the two 4-dim invariant submanifolds
of the 6-dim phase space defined by (i) $M=N$, $p_M=p_N$ (or equivalently $y=1$, $p_y=0$), and (ii) $N=R$, $p_N=p_R$
(or equivalently $y = z$, $p_y = 3 p_z$). The denomination invariant submanifolds derives from the fact that
each of them is mapped on itself by the general Hamiltonian flow (\ref{eqH}),
in other words, invariant under the flow.
We will then examine the intersection of 2-dim stable and unstable
cylinders in these two 4-dim invariant manifolds, according to the following experiments.
\par To start we fix the parameters $\kappa^2=0.5$, $\Lambda=0.001$, with
corresponding $E_{cr}=7.9096539353149939$ and $x_{cr}=15.8034567969528$. The total energy of the
system is taken $E_0=7.9096$, so that the energy available to the rotational modes will be given by
$(E_{cr}-E_0) \sim 10^{-5}$.
\par
In the first experiment we take $(x_0=x_{crit}$, $p_{x_0}=0)$, and fix initial conditions
on the 4-dim invariant submanifod $(i)$, namely, with ($y_0=1$, $p_{y_0}=0$).
That is, we are restricting ourselves to a particular domain of initial conditions
in the sector $(z, p_z)$ of the center manifold $S^3$, which has the topology of $S^1$ and is defined
by the Hamiltonian constraint (\ref{eqGen21}) as
\begin{eqnarray}
\label{eqGen4}
\nonumber
H(x=x_{cr}, p_x=0, y=1, p_y=0, z, p_z)=0.
\end{eqnarray}
By performing the evolution of orbits from a large set (of the order of $1,500$) of initial conditions in the above domain,
the exact dynamics actually evolves a 4-dim invariant subset $(x, p_x, z, p_z)$ of the full 6-dim phase space
as expected due to our restriction to the 4-dim invariant manifold $(y=1,p_y=0)$. In this particular experiment,
we have that under the exact dynamics no motion is present in the sector $(y,p_y)$.
We generate one 2-dim stable and one 2-dim unstable cylinders of orbits which initially move towards the first bounce.
\begin{figure}
\begin{center}
{\includegraphics*[height=6cm,width=9.4cm]{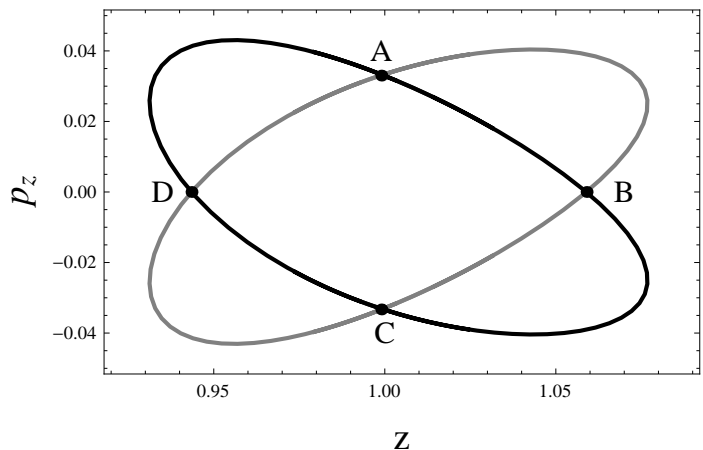}}
\caption{(first experiment) The first crossing of the stable cylinder (gray) and unstable cylinder (black)
in the surface of section $\Sigma$ (at the first bounce) shown in the plane $(z,p_z)$,
for $E_0=7.9096$. The four points $A$, $B$, $C$, $D$ are the unique points of
the first transversal crossing of the cylinders. These points define homoclinic orbits which are
contained in both unstable and stable cylinders and are bi-asymptotic to the center manifold $S^3(E_0)$,
constituting an invariant signature of chaos in the dynamics.
Here $E_{crit}=7.909653935312942$}.
\label{fig-9}
\end{center}
\end{figure}
In order to detect the first intersections of the two cylinders we adopt
$\Sigma:(x=x_b,p_x=0)$ as the surface of section, where $x_b \simeq 1.3118$
is the coordinate of the first bounce of the orbits at $p_x=0$. In Fig. \ref{fig-9}
we plot the points $(z_{b}, p_{z_b})$ of the sections of both cylinders in the first cross of $\Sigma$.
The points $A$, $B$, $C$, $D$ (contained in the sector $(z,p_z)$ of $\Sigma$) characterize the transversal
crossing of the cylinders. An detailed examination of the numerical points of the map
shows indeed that all orbits arrive
at the first bounce $(x_b\simeq 1.3118,p_x=0)$ with coordinates $y_b=1$, $p_{y_b}=0$, being a further
verification of the accuracy of our numerical treatment. The
points $A$, $B$, $C$, $D$ in Fig. \ref{fig-9} therefore define homoclinic orbits, namely,
orbits which are in the intersection of the unstable and stable cylinders, and give an invariant
characterization of chaos in the model.
These homoclinic orbits are contained both in the stable cylinder and the unstable cylinder
and are bi-asymptotic to the center manifold ${\Im}_{E_0} \subset S^3(E_0)$.
\par The coordinates $(y, p_y, z, p_z)$ of the homoclinic points
$A$, $B$, $C$, $D$ are given approximately by
\begin{eqnarray}
\label{Inters2}
\nonumber
A &\simeq& (1, 0, 0.9997818, 0.0330003),\\
\nonumber
B &\simeq& (1, 0, 1.0604379, 0.0012771),\\
\\
\nonumber
C &\simeq& (1, 0, 0.9991485, -0.0331563),\\
\nonumber
D &\simeq& (1, 0, 0.9445269, -0.0011262).
\end{eqnarray}
\begin{figure}
\begin{center}
{\includegraphics*[height=6cm,width=9.4cm]{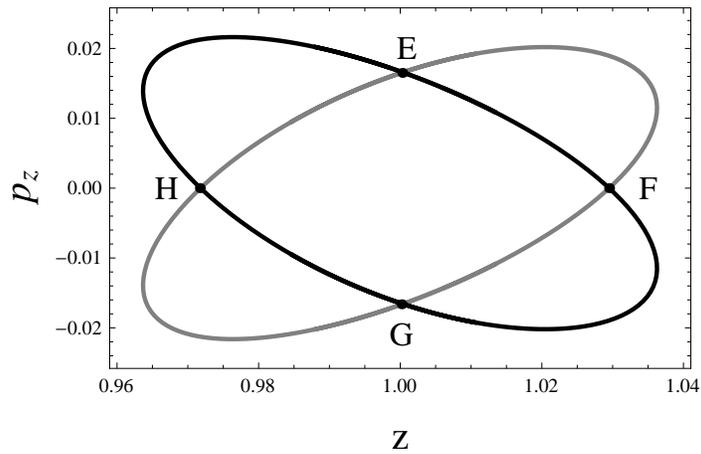}}
\caption{(second experiment) Projection on the plane $(z, p_z)$ of the first crossing of the stable cylinder (gray) and unstable cylinder (black)
in the surface of section $\Sigma$ (at the first bounce). In this projection the four points $E$, $F$, $G$,
$H$ are the unique points of the first transversal crossing of the cylinders (cf. text). These points define homoclinic
orbits which are contained in both unstable and stable cylinders and are bi-asymptotic to the center manifold $S^3(E_0)$,
constituting an invariant signature of chaos in the dynamics of the model.
}
\label{fig-10}
\end{center}
\end{figure}

\begin{figure}
\begin{center}
{\includegraphics*[height=6cm,width=9.4cm]{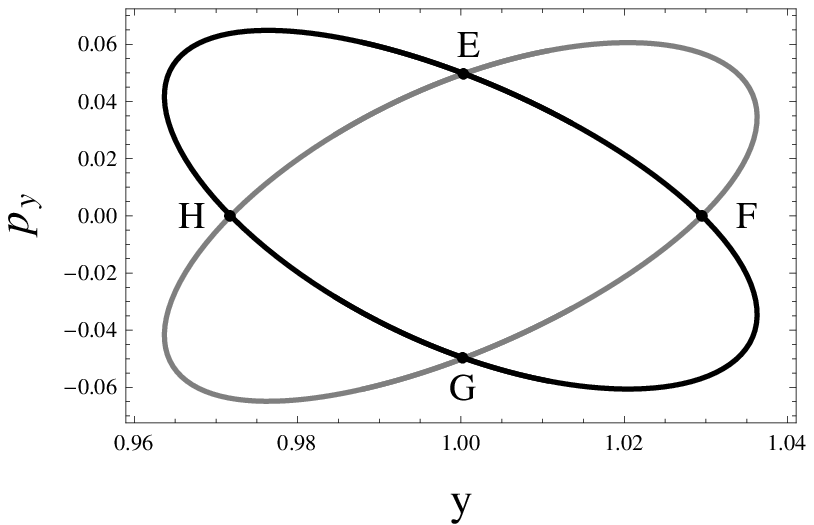}}
\caption{(second experiment) Projections on the complementary plane $(y, p_y)$ showing the first crossing of
the stable cylinder (gray) and unstable cylinder (black)
in the surface of section $\Sigma$ (at the first bounce). The four points $E$, $F$, $G$,
$H$ of figure \ref{fig-10}, characterizing the transversal crossing of the cylinders are shown, constituting
an invariant signature of chaos in the dynamics of the model.}
\label{fig-11}
\end{center}
\end{figure}
\par Analogously in the second experiment we maintain the same values for the
parameters $E_0$, $\kappa^2$ and $\Lambda$ together with the initial conditions
$(x_0=x_{crit}$, $p_{x_0}=0)$. However now fix the remaining initial conditions
on the 4-dim invariant submanifod $(ii)$ instead, namely,
$(y_0=z_0, p_{y_0}= 3 p_{z_0})$. In fact we are restricting ourselves
to a particular domain of initial conditions of the center manifold $S^3$
which has the topology of $S^1$ and is defined by the Hamiltonian constraint
\begin{eqnarray}
\label{eqGen5}
\nonumber
H(x=x_{cr}, p_x=0, y=z, p_y=3 p_z)=0.
\end{eqnarray}
With the exact dynamics we generate one 2-dim stable and one 2-dim unstable cylinders
which initially move towards the first bounce. These cylinders are generated
from a set of about $1,500$ orbits, with initial conditions taken in the above domain
which actually correspond to a flow in the 4-dim invariant submanifold $(ii)$ of the full
6-dim phase space.
\par In Fig. \ref{fig-10} we plot the points $(z_{b}, p_{z_b})$ of the sections of both cylinders
in the first crossing of $\Sigma$. The four points $E$, $F$, $G$, $H$ of the figure characterize the transversal
crossing of the cylinders, defining homoclinic orbits which are in the intersection of the unstable
and stable cylinders and are bi-asymptotic to the center manifold ${\Im}_{E_0} \subset S^3(E_0)$.
A complementary map in the sector $(y,p_y)$ is given in Fig. \ref{fig-11}, showing the first crossing of
the cylinders by surface of section $\Sigma$ at the bounce.
This map confirms the transversal crossings at the points $E$, $F$, $G$, $H$. The coordinates $(y, p_y, z, p_z)$ of
the homoclinic points $E$, $F$, $G$, $H$ are given approximately by
\begin{eqnarray}
\label{Inters2n}
\nonumber
E &\simeq& (1.0002743, 0.0496506, 1.0002743, 0.0165502),\\
\nonumber
F &\simeq& (1.0290076, -0.0013846, 1.0290076, -0.0004615),\\
\\
\nonumber
G &\simeq& (1.0002743, -0.0496506, 1.0002743, -0.0165502),\\
\nonumber
H &\simeq& (0.9721427, -0.0009974, 0.9721427, -0.0003324).
\end{eqnarray}
\par
\begin{figure}
\begin{center}
{\includegraphics*[height=6cm,width=9.4cm]{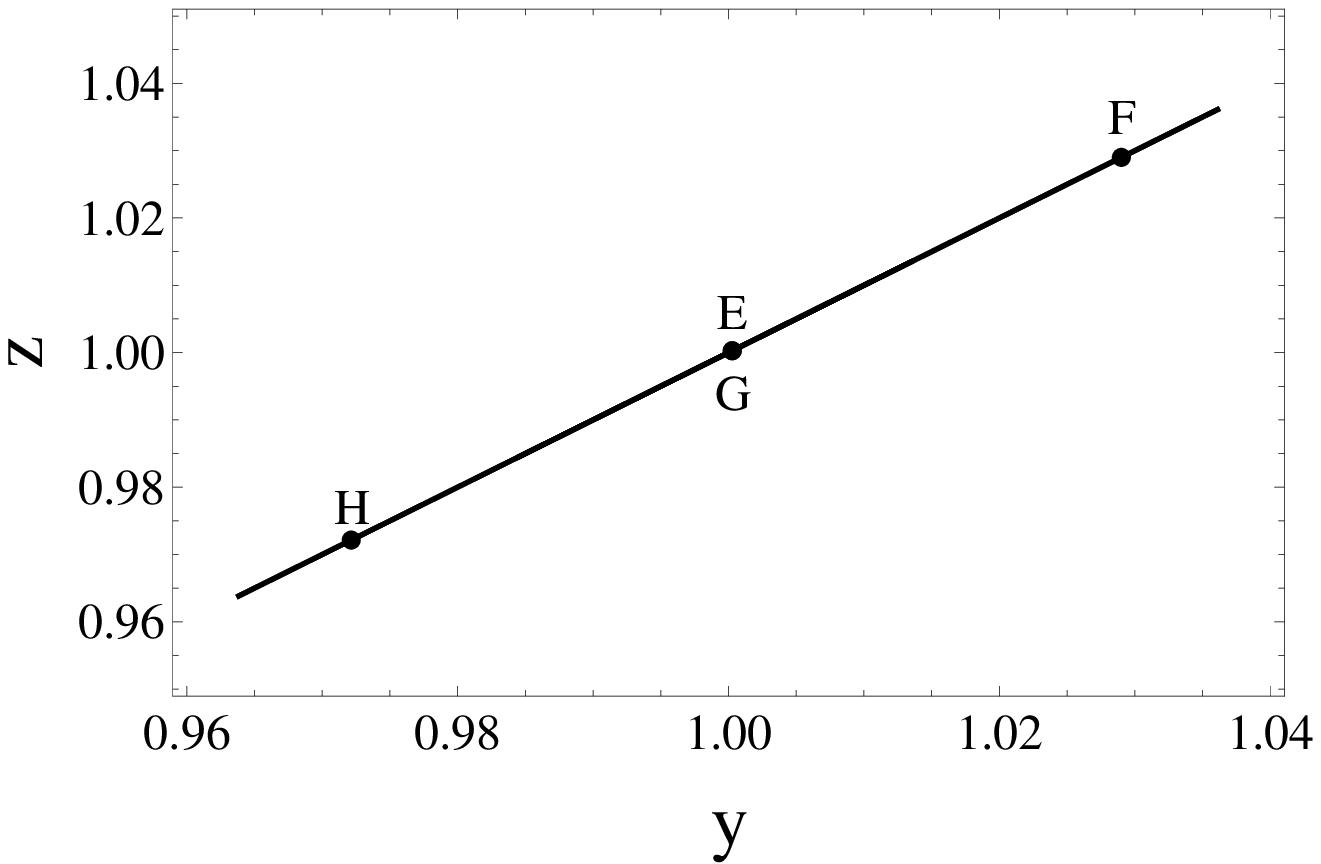}}
{\includegraphics*[height=6cm,width=9.4cm]{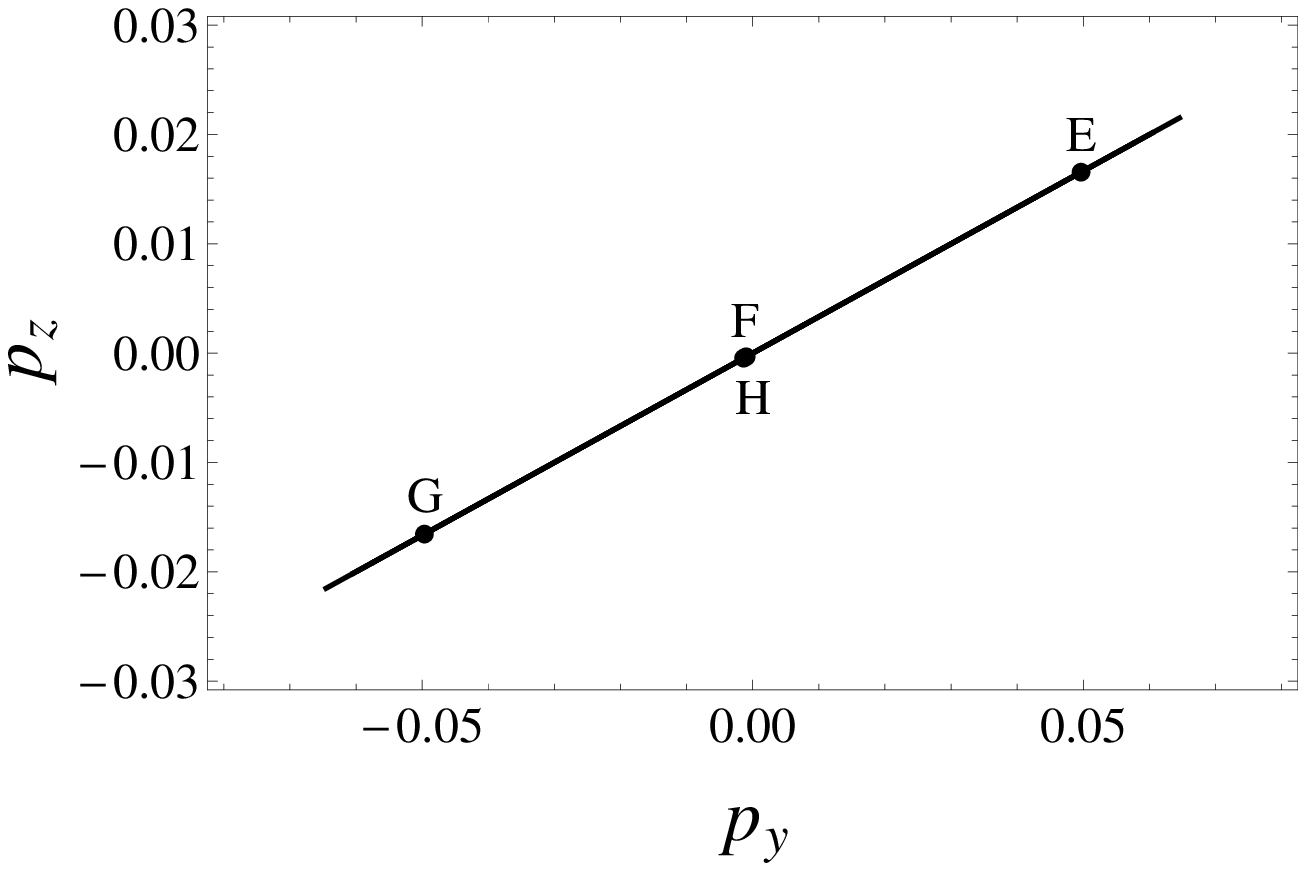}}
\caption{The first crossing of the stable and unstable cylinders with the surface of section $\Sigma$,
projected on the phase space sections $(y,z)$ and $(p_y,p_z)$. As expected
the sections of both cylinders coincide on the straight lines $y=z$ and $p_y=3p_z$. The points
corresponding to the transversal crossing of the cylinders $E$, $F$, $G$ and $H$ are plotted for reference.}
\label{figCross1}
\end{center}
\end{figure}

 An detailed examination of the numerical points of the map shows that
all orbits arrive at the first bounce $(x\simeq 1.3118,p_x=0)$ with coordinates $(y_b=z_b)$ and $(p_{y_b}=3 p_{z_b})$.
This is also illustrated in Figs. \ref{figCross1} where the first crossing of the unstable and stable
cylinders with the surface of section $\Sigma$ in the bounce are displayed. The section
of both cylinders, projected on the sectors $(y,z)$ and $(p_y,p_z)$, lie on the straight
lines $y=z$ and $p_y=3p_z$ respectively, as expected.
This is also a further verification of the accuracy of our numerical results.

\par The sets (\ref{Inters2}) and (\ref{Inters2n}) are two distinct numerical evidences of chaos in the dynamics,
and constitute an invariant signature of chaos in the model.
We must mention that the dynamics near homoclinic orbits is very complex associated with the presence of
horseshoe structures\cite{smale, moser,wiggins02,guckenheimer,ozorio2}.
The coordinates of the homoclinic points (\ref{Inters2}) and (\ref{Inters2n}) satisfy the
constraint (\ref{eqGen3}), implying that they are contained in the transversal intersection
of two $S^3$ at the bounce, namely, a $S^2$.
This fact indicates that the {\it chaotic saddle} -- connected with the structure of homoclinic orbits
bi-asymptotic to the center manifold $S^3$ -- is a Cantor-type set having the compact support $S^2$\cite{cresson}.

\section{Conclusions and final comments}

In this paper we examined the dynamics of a Bianchi IX model, with three scale factors,
sourced by a pressureless perfect fluid in the framework of bouncing Braneworld cosmology. Assuming
a timelike extra dimension and a 5-D de Sitter bulk, the modified Einstein's field equations on
the 4-dim Lorentzian brane furnish a dynamics with correction terms that
avoid the singularity and implement nonsingular bounces
in the early phase of the universe. In terms of metric functions the
modified Einstein's equations have a first integral that can be expressed
as a Hamiltonian constraint in a 6-dim phase space, yielding a three degrees of freedom
dynamical system which governs the motion in phase space.
Due to an effective cosmological constant on the brane the phase space
presents two critical points in a finite region of the phase space, a center-center-center
and a saddle-center-center, plus two critical points at infinity corresponding to the de Sitter solution.
Together with a 2-dim invariant plane of the dynamics the critical points allow to
organize the dynamics of the phase space.
\par We examine the structure of the dynamics in a linearized neighborhood of the saddle-center-center.
We identify constants of motion associated with the saddle sector,
which allow to define the linear stable $V_S$ and unstable $V_U$ manifolds. We also identify
constants of motion connected to the center-center sector, which define the center manifold of linearized
unstable periodic orbits and has the topology of $S^3$.
In the linear domain the direct product $V_S \times S^3$ and $V_U \times S^3$
define the structure of stable and unstable cylinders which constitute boundaries
in the 5-dim energy surface of the dynamics.
\par
The nonlinear extension of the center manifold of unstable periodic orbits is parametrized
by the constant of motion $E_0$ ($E_0 < E_{\rm cr}$) with the topology
of $S^3$ maintained. As one decreases the parameter $E_0$ the nonlinearity of the center manifold
increases, with a corresponding increasing of the dynamical instability as shown in our numerical simulations.
The extension of the 4-dim stable $W_S=V_S \times S^3$ and unstable $W_U=V_U \times S^3$ cylinders away from the neighborhood of the
center manifold have the structure of 4-dim spherical cylinders with the topology $R \times S^3$.
\par By a proper canonical transformation we are able to separate the three degrees of freedom of the
dynamics into one degree -- connected with the expansion and/or contraction of the scales of the model --
isolated from the other two related to pure rotational degrees of freedom associated
with the center manifold $S^3$. By expanding the Hamiltonian constraint and Hamilton's equations in these
coordinates we show that the typical dynamical flow is an oscillatory mode about the orbits of the
invariant plane. In particular the stable $W_S$ and unstable $W_U$ cylinders are composed of oscillatory orbits
about the separatrices which emerge from the saddle-center-center critical point and guide the cylinders.
These cylinders have the same energy $E_0$ of the center manifold and coalesce
to it as $t \rightarrow \pm \infty$.
As these spherical cylinders are 4-dim surfaces they separate
the 5-dim energy surface in two dynamically disconnected pieces.
This fact is a fundamental feature of the dynamics for characterization of chaos in the case of an eventual transversal intersection
of $W_S$ and $W_S$. As the separatrix which divides regions I and II in the invariant plane makes an homoclinic connection to the
saddle-center-center critical point, this fact necessarily leads to the transversal crossings of $W_S$ and $W_U$.
The transversal crossing of the cylinders consists of homoclinic orbits which are contained both in the stable and the unstable cylinder
and are biasymptotic to the center manifold $S^3$. The presence of a homoclinic orbit in the dynamics
is an invariant signature of chaos in the model\cite{smale,moser,conley,wiggins01}. The homoclinic intersection manifold
has the topology of $R \times S^2$ and consists of all homoclinic orbits biasymptotic
to the center manifold defining a {\it chaotic saddle}\cite{wiggins2} associated with $S^3$.
\par The first transversal crossings of the stable $W_S$ and unstable $W_U$ cylinders
are shown numerically in two distinct experiments. For the sake of computational simplicity we
restricted ourselves to cylinders generated from initial conditions taken on the center
manifolds of the two 4-dim invariant manifolds of the dynamics defined respectively by
$(y=1,p_y=0)$, and $(y=z,p_y=3 p_z)$. We adopted
the surface of section $\Sigma$ at the bounce defined by $(x_b,p_x=0)$ where $x_b$ is
the scale factor of the bounce for the orbits. By performing the evolution of orbits
via the 6-dim exact dynamics we generate one 2-dim stable and one 2-dim unstable cylinders of orbits,
and detected their transversal intersection corresponding to
four homoclinic points in the first crossing of $\Sigma$ by the cylinders, for both experiments.
These points define orbits which are homoclinic to the center manifold ${\Im}_{E_0} \subset S^3(E_0)$.
\par In all our numerical simulations we used the 6-dim exact dynamics, in accordance with (\ref{eqH}),
and the error in the Hamiltonian constraint (\ref{eqGen21}) is
checked to remain $ \lesssim 10^{-13}$ for all $t$.
\par We finally compare some features of the dynamics, namely
the oscillatory approach to the bounce and the chaotic behavior of
the dynamics, with analogous features present in the BKL conjecture in general relativity.
First we note that in both models the oscillatory approach to the
bounce/singularity is a key feature of the dynamics. In the general Bianchi IX
model discussed here the three degrees of freedom of the dynamics are separated into one degree
(connected with the expansion and/or contraction of the scales of the model) plus pure rotational degrees of freedom
associated with the center manifold $S^3$. The typical dynamical flow is an oscillatory
mode about the orbits of the 2-dim invariant plane; in particular from the center manifold it emerges
the stable and unstable 4-dim cylinders of oscillatory orbits that are guided towards the bounce
by the separatrix in the invariant plane. In the limit of $\kappa^2 \rightarrow 0$
(close to the general relativity dynamics) the motion
on, or about the unstable cylinder approximates the oscillatory BKL motion up to a scale $x^3 \geq \kappa^2$.
As one can make $\kappa^2$ as small as wanted, a long oscillatory approach towards a neighborhood
of $x=0$ can be developed, with a behavior analogous to one of the Kasner eras of the BKL model.
However $\kappa^2$ cannot be made equal to zero as this would correspond to a change
of topology of the phase space. The same considerations would apply for
the case of a mixmaster universe, with a nonvanishing cosmological constant and $E_0=0$, in general relativity.
Second, the chaos in the present model has a homoclinic origin, resulting from the
homoclinic transversal intersections of the stable and unstable 4-dim cylinders
emerging from the center manifold $S^3$. In contrast the chaotic behavior in the BKL dynamics
appears in a map that connects the length of the succeeding Kasner eras
in the approach to the singularity of general relativity, for which we have no counterpart.
Nevertheless, considering the general relativity limit, we have topological evidence that the 4-dim
cylinders -- emerging from the center manifold $S^3$
and guided by the separatrix connecting the saddle-center-center to the
singularity -- should intersect and generate a homoclinic orbit from this intersection.
\par In a future work we intend to examine the transversal intersection of the spherical cylinders
$R \times S^3$ in the full 6-dim phase space.
We also intend to examine the chaotic exit to the final accelerated de Sitter stage
for initial condition sets (corresponding to initially expanding universes) taken in a small neighborhood
about the separatrix $S$ approaching the saddle-center-center for $t > 0$.
As in \cite{maier}, we expect these sets
to have fractal basin boundaries connected to the code
recollapse/escape leading to a chaotic exit to
the de Sitter accelerated phase. We also expect to observe the draining of initial
condition basins from recollapse to escape behavior,
as time increases. For $t\rightarrow \infty$ only the homoclinic intersection
manifold is expected to remain in recurrent oscillatory motion.

\section*{Acknowledgements}
RM acknowledges the financial support
from CNPq/MCTI-Brasil and CAPES-FAPERJ, through a PCI-BEV research grant No. 170047/2014-8 and
Post-Doctoral grant No. 101.493/2014, respectively. IDS acknowledges the financial support
from CNPq/MCTI-Brasil, through a research grant No. 304064/2013-0. EVT acknowledges FAPES-ES-Brazil.
The figures were generated using the Wolfram Mathematica $7$.

\end{document}